\documentclass[11pt,reqno]{amsart}
\pdfoutput=1
\usepackage{packages}

\begin{document}

  \title[Quantization of the Nonlinear Sigma Model Revisited]{Quantization of the Nonlinear Sigma Model Revisited}
  \author{Timothy Nguyen}
  \address{Michign State University \\ 619 Red Cedar Road \\ East Lansing, MI 48824}
  \email{timothyn@math.msu.edu}
%  \date{\today}

\begin{abstract}
We revisit the subject of perturbatively quantizing the nonlinear sigma model in two dimensions from a rigorous, mathematical point of view. Our main contribution is to make precise the cohomological problem of eliminating potential anomalies that may arise when trying to preserve symmetries under quantization. The symmetries we consider are twofold: (i) diffeomorphism covariance for a general target manifold; (ii) a transitive group of isometries when the target manifold is a homogeneous space. We show that there are no anomalies in case (i) and that (ii) is also anomaly-free under additional assumptions on the target homogeneous space, in agreement with the work of Friedan. We carry out some explicit computations for the $O(N)$-model. Finally, we show how a suitable notion of the renormalization group establishes the Ricci flow as the one loop renormalization group flow of the nonlinear sigma model.
\end{abstract}

\maketitle

\tableofcontents

\section{Introduction}

Nonlinear sigma models provide an important class of quantum field theories, owing to the fact that they exhibit perturbative renormalizability and asymptotic freedom \cite{Fr, BLZ}, possess a rich variety of supersymmetric and Wess-Zumino-Witten extensions \cite{Wit-Bos, Wit-Hol, QFT-Math}, and describe phenomenon ranging from spontaneous symmetry breaking \cite{Meetz, Wein68} to the dynamics of string world-sheets \cite{QFT-Math, Ket}, to name a few of their many features. This paper revisits the first issue from among this list of topics via a mathematically rigorous point of view. Specifically, we address the issue of perturbative renormalizability of the nonlinear sigma model in the presence of (nonlinear) symmetries, these symmetries arising from diffeomorphism covariance and the isometric action of a Lie group in the case when the target is a homogeneous space. Such symmetries are present in the underlying classical theory and a priori can lead to anomalies whenever they are violated via the process of regularization and renormalization needed to define the quantum theory.

It has been known for a long time in the physics literature that the treatment of anomalies can be formulated as a cohomological problem \cite{BRS, LNP}. Such a formulation relies upon the many methods of perturbative renormalization developed during the 1970s and 80s. (For a general overview, see \cite{PR} and the references therein; for works specific to the nonlinear sigma model, see e.g. \cite{BLZ, BLS, BBBCD, LNP}). However, the topics discussed in such works have received considerably less attention from the mathematical community, so that navigating such literature provides a challenge to those seeking complete rigor. So as to be more specific, we would like to indicate some of the difficulties we have encountered in the footnote below.\footnote{(i) The frequently used dimensional regularization scheme \cite{BLS, BLZ} discards massless tadpoles. This can cause inconsistencies if done naively \cite{CL}. (ii) Various Ward-identities and Slavnov-Taylor identities are often invoked, but their legitimacy a priori needs to be checked against the regularization and renormalization scheme employed. It is unclear (to the author) if these checks are always done properly (in a way that holds to all orders in perturbation theory). (iii) The literature is unfortunately filled with many disclaimers concerning rigor and consistency, e.g. \cite[p. 4, 19, 148]{LNP} \cite[Ch 18.2.4]{HT} \cite{Lam}.} Despite such matters, it is not our intention here to make any definitive claims as to the legitimacy of such issues, since how one regards them is likely to involve elements of subjectivity and differing standards of evaluation. Nevertheless, we believe they strongly indicate the need for presenting a concise mathematical framework with which to understand the perturbative quantization of the nonlinear sigma model.

In this paper, we apply the manifestly rigorous formulation of perturbative quantum field theory due to K. Costello \cite{Cos} to the treatment of anomalies for the nonlinear sigma model. Here, a careful treatment of the inductive cohomological analysis of anomalies, to all orders in perturbation theory, is provided using a heat kernel regulator. Using a heat kernel regulator has the advantage that it (i) provides a natural infrared regulator without adding any additional terms in the Lagrangian to introduce a mass; (ii) is robust in that a flat space domain is not required to implement it; (iii) makes locality manifest in the construction of counterterms and the potentially anomalous terms that ensue when trying to establish Ward identities. 

To describe our results, we need a few preliminaries. First, recall that the classical action of the nonlinear sigma model is given by the functional
$$S(\pi) = \frac{1}{2}\int_{\R^2} d^2x\, g_{ij}(\pi(x))\pd_{\mu}\pi^i(x)\pd_{\mu}\pi^j(x)$$
of the field $\pi: \R^2 \to X$, where $X$ is a Riemannian manifold with metric $g_{ij}$. The  Euler-Lagrange equations of $S$ describe harmonic mappings from $\R^2$ into $X$. Next, we have to discuss how the the notion of a symmetry is expressed in the nonlinear sigma model. We are interested in those symmetries arising from the action of an isometry group and from changes of coordinates (diffeomorphism covariance). For both of these, it is much more convenient to work with infinitesimal symmetries. In the case of when we have an isometric action of a Lie group $G$, this means replacing the group action with the corresponding action of its Lie algebra $\g$. The latter action is more easily handled for the same reason that representations of Lie algebras are easier to handle than representations of Lie groups: one need not worry about whether an element of a Lie algebra can be exponentiated. Similarly, when discussing symmetries dealing with changes of coordinates, we work with its proper infinitesimal notion which is best captured in the setting of jet bundles on the target $X$. Informally, the jet bundle $\Jet(X)$ is the (infinite-dimensional) bundle over $X$ whose fiber at $p \in X$ consist of germs at $p$ of smooth functions (see Appendix \ref{SecJet}). Given a smooth function on $X$, its Taylor expansion in local coordinates about each point $p \in X$ yields a section of $\mr{Jet}(X)$. On the other hand, a section of $\Jet(X)$ arises from a smooth function only if the family of Taylor series one obtains arise from a single smooth function. More precisely, there is a connection $\nabla: \Jet(X) \to \Omega^1(\Jet(X))$ which is flat, i.e. $\nabla^2 = 0$, and those sections $s$ of the jet bundle which arise from smooth functions are precisely those which are flat: $\nabla s = 0$. As a consequence, diffeomorphism covariance is captured infinitesimally as a flatness condition (as will be explained in further detail later on). Our analysis in terms of jets is essentially the analysis used in Friedan's classic work on quantizing nonlinear sigma models \cite{Fr}.

In the quantum theory, the infinitesimal symmetries provided by the $\g$-action and the flat connection $\nabla$ will themselves receive quantum corrections. Showing that the symmetries of a classical theory are retained (in a quantum corrected form) in the quantum theory amounts to showing that the corresponding Ward identities are satisfied. In the Batalin-Vilkovisky (BV) approach to quantization \cite{BV} \cite{Wein}, this is equivalent to showing that a \textit{quantum master equation} is satisfied. In this approach, odd fields are introduced into the theory for every symmetry generator under consideration, so that symmetries of the theory can be encoded as a degree one vector field acting on the space of all fields (such fields are sections of a graded vector bundle). This degree one vector field in turn yields a degree one differential on the chain complex of local action functionals. Solving the quantum master equation involves understanding the cohomology of this differential acting on a suitable space of local action functionals. The end result is that the cohomology group in degree one parametrizes potential obstructions to solving the quantum master equation, while that in degree zero parametrizes deformations to solutions of the quantum master equation modulo equivalence.

Such a procedure is well-defined for finite-dimensional systems and is a priori ill-defined for the infinite dimensional systems pertaining to quantum field theory. One of Costello's main contributions in \cite{Cos} is to regularize and renormalize quantum field theories carefully and explicitly so that the above cohomological analysis of the quantum master equation carries over, when suitably interpreted, to all orders in perturbation theory. What the heart of our paper amounts to then, after setting up the theory properly, is an analysis of the relevant cohomology groups based on the $\g$-symmetry and the connection $\nabla$.

We obtain the following results. First, we consider the special case of $X = G/H$, for which find that there is no local (i.e. for the perturbation theory about a fixed $p \in X$) obstruction to quantizing the $\frak{g}$-symmetry. Second, we turn to the problem of global quantization (we let $p$ vary) in which case we also have to consider the cohomology of $\nabla$ on the appropriate jet bundles. In the situation where $X$ is an arbitrary Riemannian manifold, we recover Friedan's statement that there is no coordinate anomaly associated with $\nabla$. This is essentially a consequence of the fact that the de Rham complex of the jet bundle has no higher cohomology groups (Proposition \ref{PropJet}). Finally, considering $X = G/H$ globally, while there is no local anomaly for the $\g$-symmetry, there may be a global one due to the effects of monodromy (owing to the topology of $X$). These results can be summarized as follows:

\begin{Theorem}\label{MainThm}
Consider the nonlinear sigma model of maps from $\R^2$ into a Riemannian manifold $X$ with metric $g_{ij}$. Consider those quantizations which are invariant under the Euclidean group of translations, rotations, and reflections.
  \begin{enumerate}
    \item There is no obstruction to quantizing the theory globally, i.e., there is no anomaly associated to diffeomorphism covariance of the classical theory.
    \item Next, specialize to $X = G/H$ a homogeneous space with $H$ compact and assume $g_{ij}$ is $G$-invariant. Let $\mr{Met}^G$ denote the vector space of $G$-invariant elements of $\Sym^2(T^*X)$ modulo Lie derivatives of the classical metric $g_{ij}$ by $G$-invariant vector fields. Consider only those quantizations which are renormalizable.
        Furthermore, suppose that\\
        \begin{tabular}{rl}
        $\quad \mathrm{(a)}$ & $G$ is compact and semi-simple, or\\
        $\quad \mathrm{(b)}$ & $T_{[H]}X$ possesses no proper $H$-invariant subspace.
        \end{tabular}\\
        (In particular, $X$ can be an irreducible symmetric space.) Then the obstruction space associated to quantizing the classical $\g$-symmetry is $H^1(X; \mr{Met}^G)$, while the deformation space is $H^0(X; \mr{Met}^G)$. In particular, if $H^1(X) = 0$, then the $\g$-symmetry is not anomalous.
  \end{enumerate}
\end{Theorem}

The statement about the deformation space above means that order by order in perturbation theory (i.e. in the perturbative parameter $\hbar$), the space of $\g$-invariant quantizations is isomorphic to $H^0(X; \mr{Met}^G)$, corresponding to the freedom in the choice of renormalization scheme at that order in perturbation theory. Observe that if $H$ acts irreducibly on the tangent space of $[H] \in G/H$ (e.g. if $X$ is an irreducible symmetric space) then $\mr{Met}^G$ is just the space of $G$-invariant metrics. Thus, the space of quantizations is given by a single renormalized coupling constant, a $G$-invariant metric valued in formal power series in $\hbar$ which reduces to the classical metric modulo $\hbar$. The hypothesis of renormalizability is a hypothesis about the scaling law of the effective interactions one obtains. Rather than defining what this means in generality, in the presence of the $G$-symmetry above, what it essentially amounts to is a scaling symmetry of the theory (see p. \pageref{renorm}). A more thorough discussion of renormalizibility can be found in \cite{Cos}. Finally, it is unclear if hypotheses (a) and (b) are essential; we only used them as a simplifiying assumption in the computation of cohomology groups. Our results are in very similar agreement with the work of Friedan \cite[p. 382]{Fr}.

At the end of this paper we also explain how there is a natural notion of a renormalization group acting on the space of quantizations of the nonlinear sigma model, out of which we produce a rigorous derivation of Friedan's famous result:

\begin{Theorem}\label{MainThm2}
The one-loop renormalization group flow for the nonlinear sigma model equals the Ricci flow on the target manifold.
\end{Theorem}

The use of rigor in proving this theorem lies chiefly in showing that there is a globally consistent quantization, in the sense that there is no diffeomorphism covariance anomaly, which is a consequence of Theorem 1. This procedure in Friedan's work \cite{Fr} was done in Section 6.3, which in addition to being subject to the caveats of the general literature discussed above, has a very noticeable missing step on p. 371 where it makes reference to a ``standard argument by induction" (to show that there is no cohomological obstruction to eliminating a potential anomaly). This argument is precisely the cohomological analysis we carry out in this paper, which to the best of our knowledge, does not appear elsewhere in the literature. Once it is shown that there is no anomaly, formulating the scaling action that defines the renormalization group and showing that the one-loop beta function is the Ricci tensor is, by now, a routine procedure. For a more detailed analytic approach to the renormalization group in the setting of nonlinear sigma models, see e.g. \cite{GK2, MR}.

We should remark that another approach to the issue of diffeomorphism covariance in the nonlinear sigma model involves using the (covariant) background field method \cite{Abb, Hon, AGM}, which among other things, involves developing covariant methods of renormalization. For this latter step, one wants counterterms to be formed out of tensor fields on the target of the nonlinear sigma model (as opposed to being given by coordinate-dependent expressions). However, as discussed in \cite{AGM, Stelle}, there will also be ``off-shell" counterterms arising from reparametrizing the theory by an infinitesimal diffeomorphism (for an explicit example of this in the $O(n)$-model, see the end of Section \ref{SecON}). It is unclear to the author if the background field method really addresses (or is the right language for) the issue of diffeomorphism covariance, since as can be seen in our work (and that of Friedan \cite{Fr}), the flat connection $\nabla$ encoding diffeomorphism covariance itself receives quantum corrections due to renormalization. The author is unable to see how this issue is naturally taken into account using the background field method. Nevertheless, the background field method seems to be a standard ``quick and dirty way" of analyzing beta functions of nonlinear sigma models, in particular, showing that it is equal to the Ricci tensor at one loop.

Note that in our analysis we always work with a fixed infrared regulator (the parameter $L$ in our heat kernel regulator). This is to avoid the difficulties involved with infrared problems for massless theories in two dimensions. On the other hand, it is known that eliminating infrared divergences may lead to additional obstructions to preserving a symmetry at the quantum level (we have treated the latter as a purely ultraviolet problem). See e.g. \cite{BPS, CPS}.

Finally, we note that it has been indicated to us by the referee that work in a very similar spirit to ours on coset models has appeared in \cite{BBBCD}. Moreover, removing the infrared regulator is addressed there.

The outline of this paper is as follows. In Section 2, we provide the setup for Batalin-Vilkovisky geometry in terms of odd symplectic geometry and then set up the nonlinear sigma model in that framework. In Section 3, we explain our heat kernel regularization and renormalization procedure involved for quantization. In Section 4, we perform cohomological computations for the local quantization of the $\g$-symmetry. In Section 5, we provide an illustration of our analysis through the well-studied $O(N)$-model. In Section 6, we study the global quantization problem and prove our main theorems. The appendix contains background material on the tools and notation we use, namely those arising in the context of graded manifolds, Lie algebra cohomology, and jet bundles. To make our paper as self-contained as possible, we also give a brief exposition of Wick's Theorem, mostly for notational purposes, and we motivate the definitions for the effective interactions and the regulated quantum master that we use.

We note that there is another mathematically rigorous approach to perturbative quantization based upon the Batalin-Vilkovisky formalism in Lorentzian signature due to K. Rejzner and K. Fredenhagen \cite{R, FR}. While the literature on rigorous mathematical aspects of quantum field theory is of course very vast, \cite{Cos} and \cite{R} are the only works of which we are aware that develop the Batalin-Vilkovisky formalism in a transparently rigorous manner.\\

\noindent\textit{Acknowledgements.} This paper could not have been written without the significant guidance provided by Kevin Costello in patiently and generously explaining his approach to quantum field theory to the author. Kevin also deserves credit for helping simplify some of the arguments in this paper. The author would also like to thank Si Li for explaining some of the details of his paper \cite{LiLi}, which was then adapted and incorporated into Section \ref{SecGlobal}, and Malek Abdesselam for pointing out some references. Finally, the author would like to thank Michael Douglas, Ryan Grady, Owen Gwilliam, Martin Rocek, Robert Shrock, and his many colleagues at the Simons Center for valuable discussions.

\section{Batalin-Vilkovisky Geometry}

\label{SecBV}

We set up the necessary geometric background in order to perform quantization in the Batalin-Vilkovisky (BV) formalism. Our presentation will be rather compressed and adapted to our specific needs; for additional details, see \cite{AKSZ}, \cite{Cos}, \cite{Sch}. The reader may wish to jump ahead to Section \ref{Sec2.1} or work with the concrete example in Section \ref{SecON} to balance the abstract presentation which follows.

We begin with the finite dimensional situation. Let $M$ be a smooth (connected) manifold. In what follows, we equip $M$ with a sheaf of graded algebras and perform constructions in the world of graded manifolds (see Appendix \ref{SecGM} for background). The graded manifold in question is the shifted cotangent bundle $T^*[-1]M$, whose underlying manifold is $M$ and whose sheaf of algebras is
$\cO(T^*[-1]M) = \Lambda^{-*}(TM),$
the sheaf of multi-vector fields with grading reversed. That is, the $i$th exterior power $\Lambda^i(TM)$ is placed in degree $-i$, i.e., it is the space $\Sym^i(T[1]M)$ as sheaves of graded vector bundles.

The graded manifold $T^*[-1]M$ comes equipped with a symplectic form of degree $-1$ induced from the canonical symplectic form on $T^*M$. We can define a degree one Poisson bracket $\{\cdot,\cdot\}$ on $\cO(T^*[-1]M)$ from the symplectic form on $T^*[-1]M$. In local Darboux coordinates $(x^i,\xi_i)$, where $x^i$ are even (i.e. ordinary) coordinates on $M$ and $\xi_i$ are odd coordinates corresponding to the components of cotangent vectors with respect to the basis $dx^i$ for $T^*M$, we choose sign conventions so that the Poisson bracket is given by the formula
$$\{F,G\} = \pd_{\xi_i}F\pd_{x^i}G + (-1)^{|F|}\pd_{x^i}F \pd_{\xi_i}G.$$
The Poisson bracket is thus (up to sign convention) the Schouten-Nijenhuis bracket on multivector fields. It satisfies
\begin{align*}
  \{V,f\} &= V(f) \\
  \{U,V\} &= [U,V],
\end{align*}
for $f$ any function on $M$ and $U, V \in T[1]M$ vector fields. The bracket extends to the rest of $\Sym^*(T[1]M)$ by the Poisson identity
\begin{equation}
  \{F,GH\} = \{F,G\}H + (-1)^{|G|(|F|+1)}G\{F,H\}. \label{eq:poisson-id}
\end{equation}

With the above setup, the only functions of degree zero on $T^*[-1]M$ are the ordinary functions on $M$. If we are given an action of a Lie group $G$ on $M$, and hence an induced action of $\g = \mr{Lie}(G)$ on $\cO(T^*[-1]M)$, we can extend the graded manifold $M$ in a way that allows for nontrivial functions of degree zero as follows.

Consider the vector space $\g[1] \oplus \g^*[-2]$. It is also canonically a symplectic vector space with symplectic pairing of degree $-1$. (This reduces to the previous example if we allow $M$ to have odd directions and take $M = \g[1]$.) There is a canonical function $S_\g$ on this space such that $\{S_g,\cdot\}$ squares to zero and for which the chain complex $\Sym\Big((\g[1] \oplus \g^*[-2])^*, \{S_g,\cdot\}\Big)$ is precisely $C^*(\g, \Sym(\g[2]))$, the Chevalley-Eilenberg cochain complex of $\g$ with coefficients in $\Sym(\g[2])$ (see Appendix \ref{sec:CE}). Here, the action of $\g$ on $\Sym(\g[2])$ is the one given by the adjoint action. Explicitly, if $e_a$ denotes a basis for $\g$, with $[e_a,e_b] = T^c_{ab}e_c$, then if we let $\omega_a \in \g[1]^*$ and $\omega_a^\vee \in (\g^*[-2])^* = \g[2]$ denote corresponding dual coordinates on $\g[1]$ and $\g^*[-2]$, respectively, then
\begin{equation}
S_\g = \frac{1}{2}\omega^\vee_a T^a_{bc}\omega_b\omega_c. \label{eq:Sg}
\end{equation}
The elements of $\g[1]$ and $\g^*[-2]$ are referred to as \textit{ghost} and \textit{anti-ghost} fields, respectively.

Endow $T^*[-1]M$ with the constant sheaf of vector spaces $\g[1] \oplus \g^*[-2]$. In this way, the corresponding space of functions we obtain is
\begin{equation}
  \cO\Big(T^*[-1]M \oplus \g[1] \oplus \g^*[-2]\Big) = \cO(T^*[-1]M) \otimes_\R C^*(\g, \Sym(\g[2])) \label{BVfun}
\end{equation}
as sheaves of graded $C^\infty(M)$-algebras. Given any element of (\ref{BVfun}), its polynomial degree in $\g[1]$ is referred to as the \textit{ghost degree}. The Lie group action of $G$ on $M$ induces a Lie algebra homomorphism $\rho: \g \to \Gamma(TM)$ given by
$$\rho(Z)_p = -\frac{d}{dt}\bigg|_{t=0}e^{tZ}\cdot p, \qquad p \in M.$$
Shifting degrees, we get a map $\rho: \g[1] \to \Gamma(T[1]M)$ and thus a degree zero element $S_\rho$ of $(\g[1])^* \otimes \Gamma(T[1]M)$. Observe that Poisson bracket with $S_\rho + S_\g$ yields the Chevalley-Eilenberg differential for $C^*(\g, \Sym(\g[2]) \oplus \cO(T^*[-1]M))$.

Finally, let $S_0$ denote a function on $M$. Suppose it is $\g$-invariant. Then if we let $S = S_0 + S_\rho + S_\g$, then $S$ is an element of (\ref{BVfun}) of degree zero and satisfies the classical master equation
$$\{S,S\} = 0.$$
This equation captures all the following identities by grouping the terms of $\{S,S\}$ by ghost number (polynomial degree in $\g[1]$):
\begin{equation}
  \begin{cases}
  \textrm{degree one: $\{S_\rho, S_0\} = 0$. This says $S_0$ is $\g$-invariant;} \\
  \textrm{degree two: $\frac{1}{2}\{S_\rho,S_\rho\} + \{S_\g, S_\rho\} = 0$. This says $\g$ acts as a Lie algebra on $\cO(T^*[-1]M)$;} \\
  \textrm{degree three: $\{S_\g, S_\g\} = 0$. This says the Lie bracket on $\g$ satisfies the Jacobi identity.}
\end{cases} \label{CMEprop}
\end{equation}

To summarize, we have shown the following:

\begin{Lemma}
Any function defined on $M$ which is invariant with respect to a group action $G$ on $M$ can be expressed in a canonical way as a solution of the classical master equation on the graded manifold $T^*[-1]M \oplus \g[1] \oplus \g^*[-2]$.
\end{Lemma}

Let $d\mu$ be a density on $M$. In classical mechanics, one is interested in the critical points of $S_0$ whereas in the quantum theory, one is interested in the integral of $e^{S_0/\hbar}d\mu$ over $M$. Consequently, the $\g$-invariance of $S_0$, a symmetry of the classical theory, becomes a symmetry of the quantum theory only if the measure $e^{S_0/\hbar}d\mu$ is $\g$-invariant. That is, we must have
\begin{equation}
\{S_\rho, S_0\} + \hbar\div_{d\mu}S_\rho = 0, \label{QMErho}
\end{equation}
where given a vector field $Z$, $\div_{d\mu}Z$ computes the divergence of $Z$ with respect to $d\mu$, i.e., $\L_X d\mu = \mr{div}(X) d\mu$, where $\L_X$ is the Lie derivative. The term $\div_{d\mu}S_\rho$ is defined by applying the divergence operator to each of the vector fields appearing in $S_\rho$ arising from the $\g$-action. Setting the operator $\div_{d\mu}$ to be zero on the other terms $S_0$ and $S_\g$ in $S$, we can encode (\ref{QMErho}) and the last two equations in (\ref{CMEprop}) via the \textit{quantum master equation}:
\begin{equation}
  \frac{1}{2}\{S,S\} + \hbar \div_{d\mu} S = 0. \label{QME-finite}
\end{equation}
So altogether, (\ref{QME-finite}) is an $\hbar$-deformation of the classical master equation (which in the present case differs from the classical master equation only in ghost degree one).

For the purposes of perturbative quantum field theory, we will need a version of the above results that is adapted to formal functions, i.e., those defined as a formal power series. Thus, let $\widehat{\cO}_m(T^*[-1]M)$ denote the ring of formal power series functions on $T^*[-1]M$ centered at $m \in M$ (our notation follows the algebro-geometric notion of a ring completion). While the $G$ action on $M$ no longer induces an action on $\widehat{\cO}_m(T^*[-1]M)$, we do have an action of $\g$ on $\widehat{\cO}_m(T^*[-1]M)$, since $Z \in \g$ acts via Lie differentiation on formal multivector fields.

It follows that given a degree zero element $S_0 \in \widehat{\cO}_m(T^*[-1]M)$ that is $\g$-invariant, we obtain a corresponding solution $S$ of a classical master equation in exactly the same way as before. We thus have

\begin{Lemma}\label{LemmaBV}
  Any function defined as a formal power series about a point $m \in M$ and which is $\g$-invariant can be expressed in a canonical way as an element of $C^*(\g, \Sym(\g[2]) \oplus \widehat{\cO}_m(T^*[-1]M))$ that solves the classical master equation.
\end{Lemma}

In what follows, we will recast the nonlinear sigma model as a solution of the classical master equation and then make precise sense of the corresponding quantum master equation in Section \ref{SecRR}.

\subsection{The Nonlinear Sigma Model in the Classical BV formalism}
\label{Sec2.1}

We are interested in studying the nonlinear sigma model perturbatively about constant maps from $\Sigma=\R^2$ to the target manifold $X$. Indeed, in the semiclassical limit $\hbar \to 0$, formally the partition function
$$Z = \int D\pi e^{-S(\pi)/\hbar}$$
localizes on the lowest energy configurations of $S$. Fix a constant map, i.e. a point $p \in X$. A choice of local coordinates centered at $p$ yields a diffeomorphism from a neighborhood of the origin of $T_pX$ to a neighborhood of $p$ in $X$, so that under such a correspondence, objects defined on $X$ (near $p$) can be pulled back to (a subset of) $T_pX$. In particular, elements of $\mr{Maps}(\Sigma, X)$ taking values near $p$ correspond to elements of $\mr{Maps}(\Sigma, T_pX)$ taking values near the origin. Such a precise correspondence, however, is not what is needed to formulate perturbation theory for the nonlinear sigma model. Because perturbation theory allows one to work with quantities defined at the level of formal power series, we can expand our action functional as a power series in the linearized fields $\mr{Maps}(\Sigma, T_pX)$. For the nonlinear sigma model, this is achieved by expanding the Riemannian metric in a Taylor series about the origin in $T_pX$ using the coordinate system at $p$, and then evaluating this power series on an element of $\mr{Maps}(\Sigma, T_pX)$. Such formal power series functionals are unproblematic for the usual setting of perturbative quantum field theory, since for instance, any fixed $n$-point correlation function only involves Feynman diagrams with $n$ external tails and thus makes sense as a formal power series in the perturbative parameter $\hbar$. Choosing a family of coordinate systems for every point of $X$, we obtain in this way, a family of classical field theories for every $p$ defined in terms of power series functionals.

Our next step is to introduce symmetries for the perturbatively defined nonlinear sigma model and recast this data in terms of the general framework of the BV formalism. For a general target $X$, when we linearize our theory about $p \in X$, the space of ordinary (bosonic) fields becomes $\mr{Maps}(\Sigma, T_pX)$ as discussed above. The BV formalism requires that we introduce \textit{antifields} which in the present case are elements of $\mr{Maps}(\Sigma, T_p^*X[-1])$. These odd anti-fields serve as sources to which we can couple vector fields corresponding to symmetries. When $X = G/H$ is a homogeneous space, we also include in our space of fields the space of ghosts $\g[1]$ and anti-ghost $\g^*[-2]$ to encode the global $G$-action on $X$ infinitesimally.

Altogether, the space of fields for the nonlinear sigma model linearized about $p \in X$ is
$$\cE_p = \left(C^\infty(\Sigma) \otimes_\R (T_pX \oplus T^*_pX[-1])\right) \oplus (\g[1] \oplus \g^*[-2]),$$
where we take $\g = 0$ if $X$ is not a homogeneous space. That is, the space of non-ghost fields is the space of sections of the trivial bundle $E = E(p)$ over $\Sigma$ with fiber $T_pX \oplus T^*_pX[-1]$. On the fibers $T_pX \oplus T^*_pX[-1]$ we have both the canonical odd symplectic pairing and symmetric pairing obtained from the natural evaluation of $T^*_pX$ with $T_pX$. We let
\begin{align*}
  \pi^i & \in \Gamma(\Sigma; T_pX) \\
  \check\pi^i & \in \Gamma(\Sigma; T_p^*X[-1])
\end{align*}
denote the ordinary field and anti-field variables, respectively. 

Local action functionals are given by evaluating (a power series) polydifferential function of the fields and integrating against a density (see Appendix \ref{SecJet}). When we have ghost and anti-ghost fields, we also allow our local action functionals to be polynomial in them. Finally, local action functionals are only defined modulo a Lagrangian density which is exact. Thus, we have
\begin{equation}
  \cO_\loc(\cE_p) = \mr{Dens}_\Sigma \otimes_{D_\Sigma} \mr{PolyDiff}(E) \otimes C^*(\g, \Sym(\g[2])), \label{eq:localO}
\end{equation}
where $\otimes_{D_\Sigma}$, which denotes tensor product in the category of $D_\Sigma$-modules, captures the fact that total derivatives can be, via formal integration by parts, shifted from densities to polydifferential expressions and vice versa (this implements equating Lagrangian densities that differ by an exact term).

A choice of coordinate system at $p$ means that the Taylor series of any tensor on $X$ at $p$ yields a polydifferential function on the space of fields $\cE_p$ at $p$. Paired with a density on $\Sigma$, this yields a local action functional. Let us spell this out more explicitly. It suffices to consider covariant and contravariant tensors on $X$ separately. Let $y^i$ denote the coordinates on $T_pX$ determined by our chosen coordinate system, $i = 1,\ldots, n = \dim(X)$, and let $x^\mu$ denote the coordinates on $\Sigma$, $\mu = 1,2$. Given a function $f$ on $X$, its Taylor expansion with respect to the coordinate system $y^i$ is $f_I(p)y^I/I!$, where $I = (i_1,\ldots, i_n)$ is a multi-index, $f_I(p) = \pd^{i_1}_{y^1}\cdots\pd^{i_n}_{y^n}f(p)$, $I! = i_1!\cdots i_n!$, and the sum over $I$ is implicit. We thus obtain a translation-invariant power series function of the bosonic field by evaluating the Taylor series components on the linearized field $\pi^i: \Sigma \to T_pX$. Integration against the unique (up to scaling) translation-invariant volume form $d^2x$ on $\Sigma$ yields our corresponding local action functional
$$S_p[f](\pi^i) := \int_\Sigma d^2x \frac{f_I(p)}{I!}\pi^I(x).$$

We can reformulate the above procedure using the language of jet bundles (see Appendix \ref{SecJet}) which will be useful later when performing cohomological computations. Namely, via a coordinate system, we can make the identification
\begin{equation}
\Jet_p(X) \cong \wSym(T^*_pX). \label{Jet=Sym}
\end{equation}
The above construction which associates an element of $\Jet_p(X)$ to  a local action functional is thus the statement that $\Jet_p(X)$ is isomorphic to the space of functions on a formal neighborhood of the origin in $T_pX$. Since $\pi^i$ is valued in $T_pX$, then by pointwise evaluation, any element of $\wSym(T^*_pX)$ yields a constant section of $\wSym(E^*)$.

This analysis generalizes to higher rank tensors. For tensors $T$ belonging to $\Sym^k(T[1]X)$, we proceed as before, since $\mr{jet}_p(T) \in \wSym(T^*_pX) \otimes_\R \Sym^k(T_pX[1])$, and we can evaluate the $\wSym(T^*_pX)$ factor on the bosonic field $\pi^i$ and the $\Sym^k(T_pX[1])$ factor on the anti-fields to get a $k$-linear function of the $\check{\pi}^i$. Explicitly, if $\jet_p(T) = \left(\frac{T^{i_1\ldots i_k}_I(p)}{I!} \pd_{y^{i_1}}\wedge \cdots \wedge \pd_{y^{i_k}}\right)_I$, then
$$S_p[T](\pi^i, \check\pi^i) = \int d^2x \frac{T^{i_1\ldots i_k}_I(p)}{I!}\pi^I(x) \check{\pi}^{i_1}(x)\cdots \check{\pi}^{i_k}(x).$$

For tensors $T$ belonging to $\Sym^k(T^*X)$, in particular $T$ the metric tensor belonging to $\Sym^2(T^*X)$, we have the following. First, for $k = 1$, observe that the pullback of a coordinate $1$-form on the vector space $T_pX$ under $\pi^i: \Sigma \to T_pX$ is a $1$-form on $\Sigma$ valued in coordinates on the $1$-jet of $\pi^i$, i.e., it is an element of
$\Lambda^1(\Sigma) \otimes_{C^\infty(\Sigma)}\Jet(E)^*$. Indeed, if $dy^i$ are the basis coordinate one-forms on $T_pX$, its pullback under $\pi^i: \Sigma \to T_pX$ is the one form which maps $\pd_\mu$ to $\pd_\mu \pi^i$.  Thus, for a general jet of a tensor $\left(\frac{T_I^{i_1\ldots i_k}(p)}{I!} dy^{i_1}\cdots dy^{i_k}\right)_I \in \wSym(T_p^*X) \otimes \Sym^k(T_p^*X)$, the local functional of $\pi^i$ we obtain is of the form
$$\wSym(E^*) \otimes_{C^\infty(\Sigma)} \Sym^k_{C^\infty(\Sigma)}\left(\Lambda^1(\Sigma) \otimes_{C^\infty(\Sigma)} \Jet(E)^*\right) \subset \mr{PolyDiff}(E).$$
In particular, for $k = 2$, we can contract the $\Sym^2(\Lambda^1(\Sigma))$ factor with the constant two-vector field $\pd_\mu\pd_\mu$ on $\Sigma$ and then multiply with $d^2x$ to obtain a density. Thus, this defines for us the local action functional for a general two tensor, in particular, the Riemannian metric $g_{ij}$ on $X$:
\begin{equation}
  S_p[g_{ij}] = \frac{1}{2}\int_\Sigma d^2x \frac{g_{ij,I}(p)}{I!}\pi^I(x) \pd_\mu\pi^i(x)\pd_\mu\pi^j(x). \label{metric-action}
\end{equation}
Here, we insert the factor of $1/2$ as a convenient choice of normalization.

Altogether, we get a total action functional $S_p$ for the nonlinear sigma model linearized at $p$:
\begin{equation}
  S_p = S_p[g_{ij}] + S_p[\rho] + S_\g \label{eq:Sp}
\end{equation}
Here, $S_p[\rho]$ is applied $\g$-linearly to the tangent vectors encoded by $\rho \in \g[1]^* \otimes \Gamma(TX)$. The term $S_\g$ is  defined  as in (\ref{eq:Sg}). Fixing $p$ for the time being, we drop it from the notation and write $S = S_p$. We have that $S$ naturally splits as a sum $S = S_{\mr{kin}} - I$, consisting of a kinetic part
\begin{equation}
S_{\mr{kin}} = \frac{1}{2}\int_\Sigma d^2x g_{ij,0}(p)\pd_\mu\pi^i(x)\pd_\mu\pi^j(x) \label{Skin}
\end{equation}
and a remaining interaction part $I$ given by
\begin{equation}
  I = -(S - S_{\mr{kin}}). \label{interaction}
\end{equation}
The interaction consists of the terms which are at least cubic coming from the metric and those which arise from $S_\rho$.

Next, we define the Poisson bracket on local action functionals. The symplectic pairing on the fibers of $E$, together with our choice of density $d^2x$ on $\Sigma$, allows us to define the Hamiltonian vector field of any local action functional. Namely, if $S$ is any local action functional, we have the variational derivative (see Appendix \ref{SecJet})
$$\delta S \in \Dens(\Sigma) \otimes_{C^\infty(\Sigma)} (E^* \otimes_{C^\infty(\Sigma)} \mr{PolyDiff}(E))$$
The nodegenerate symplectic pairing on the fibers of $E$ yields an isomorphism $E \cong E^*$ and and the nowhere vanishing density $d^2x$ on $\Sigma$ yields an isomorphism $\mr{Dens}(\Sigma) \cong C^\infty(\Sigma)$. Together, this yields the bundle isomorphism $\Dens(\Sigma) \otimes_{C^\infty(\Sigma)}E^* \cong E$, so that we obtain from $\delta S$, the local vector field
$$X_S \in  \wSym(\Jet(E)^*) \otimes_{C^\infty(\Sigma)}  E.$$
If $S'$ is any other local action functional on $\cE$, then we can contract $X_S$ with $\delta S'$ via the natural pairing
$$E \otimes E^* \to C^\infty(\Sigma)$$
which extends $\wSym(\Jet(E^*))$-linearly. We denote this pairing by $X_S(S')$.

\begin{Definition}\label{def:PB}
Let $S,S'$ be two local action functionals belonging to $\cO_\loc(\cE_p)$.
  (i) Suppose $S$ and $S'$ are independent of $\g$. Then the Poisson bracket $\{S,S'\}$ is the local action functional defined by
  $$\{S,S'\} = X_S(S').$$
  (ii) The Poisson bracket extends to functionals depending on $\g$ by combining the previous Poisson bracket with the Poisson bracket on $\g[1] \oplus \g^*[-2]$ using the Poisson identity (\ref{eq:poisson-id}).
\end{Definition}
The definition of the Poisson bracket extends straightforwardly to those $S'$ which are not local but which are given by polydifferential functions on the external tensor product $\boxtimes^k E$ defined over $\Sigma^k$. We must consider these latter types of functionals since they are precisely what we obtain from Feynman diagrams constructed out of local action functionals.

Having defined the Poisson bracket, we can now encode the symmetries of the nonlinear sigma model, linearized about $p$, using the classical master equation. Namely, for every $p$, we have
\begin{equation}
\{S_p,S_p\} = 0.
\end{equation}
This expresses that (i) $\jet_p(g_{ij})$ is $\g$-invariant; (ii) the map $\rho: \g \to \Jet_p(T[1]X)$ is a Lie algebra homomorphism (iii) the Lie bracket on $\g$ satisfies the Jacobi identity.

We have thus captured the $\g$-symmetry of the nonlinear sigma model linearized about a point $p$ in terms  of a classical master equation. It remains to do the same for the diffeomorphism covariance of the global nonlinear sigma model, that is, for $p$ varying. However, defining the classical master equation globally requires a bit of setup which we address in Section \ref{SecGlobal}. For now, we adapt some of the previous analysis to the global situation to set the foundation for what will come later.

For general Riemannian target $X$, the above construction yields a family of local action functionals $S_p$ determined from the section $\jet(g_{ij})$ of $\Jet(\Sym^2(T^*X))$. Recall that an arbitrary section of $\Jet(\Sym^2(T^*X))$ has independent Taylor series components at each point $p$ while those arising from the jet of a globally defined metric $g_{ij}$ have compatibly related Taylor series as $p$ varies, i.e. $\jet(g_{ij})$ is a flat section with respect to the natural connection $\nabla$ on $\Jet(\Sym^2(T^*X))$. This is to be interpreted as a diffeomorphism covariance of the underlying classical theory, since it expresses how the Taylor series of $g_{ij}$ transform under a change of coordinates. As observed by Friedan \cite{Fr}, a priori, the quantum theory, being a perturbative quantization at each point of $p$, consists of an independent renormalized metric for every $p$. What one wants to show is that there should be a single renormalized metric from which the family of renormalized metrics are obtained in a compatible way. This would thus be a preservation of the underlying classical diffeomorphism covariance at the quantum level.

In more detail, we have the following setup. Let
$$\cE = C^\infty(\Sigma) \otimes_\R \Gamma(TX \oplus T^*X[-1]) \oplus (\g[1] \oplus \g^*[-2]).$$
be the space of fields that map into an arbitrary tangent space of $X$. Equivalently, observe that the non-ghost fields of $\cE$ are sections of the bundle $\tilde E$ over $\Sigma \times X$ obtained by pullback of $TX \oplus T^*[-1]X$ under the projection $\Sigma \times X \to X$.

Local action functionals on the space of global fields are $\Omega^*(X)$-linear elements formed out of polydifferential functions of the $\cE$: \begin{align}
  \cO_\loc(\cE) &= \mr{Dens}_\Sigma \otimes_{D_\Sigma} \mr{PolyDiff}_{\Omega^*(X)}(\tilde E, \Omega^*(X)) \otimes C^*(\g, \Sym^*(\g[2])). \label{eq:globalO}
\end{align}
Indeed, since connections map bundles to bundles tensored with forms on $X$, our space of action functionals must include the de-Rham forms on $X$. Here, $\mr{PolyDiff}_{\Omega^*(X)}(\tilde E, \Omega^*(X))$ denotes those polydifferential operators which are $\Omega^*(X)$-linear. That action functionals are $\Omega^*(X)$-linear just means they are parametrized by the basepoints $p$ of $X$,  e.g., (\ref{metric-action}), regarded as a function of the basepoint $p$, is linear with respect to multiplication by functions of $p$, or more generally differential forms on $X$. One way to interpret the definition of $\cO_\loc(\cE)$ is that one has an $\Omega^*(X)$-sheaf of local action functionals over $X$, where we can restrict such functionals to a point $p$ in $X$ to obtain the space $\cO_\loc(\cE_p)$.

The flat connection $\nabla$ on the jet bundle yields a differential on $\cO_\loc(\cE)$. Indeed, $\nabla$ acts $\g$- and $D_\Sigma$-linearly and our family of normal coordinate systems allows us to transfer the connection $\nabla$ on $\Jet(X)$ and $\Jet(T[1]X)$ to $\wSym(T^*X)$ and $\wSym(T^*X)\otimes \Sym(T[1]X)$ inside the space of local action functionals. The action functional $S_p$ with $p$ varying gives us a global action functional $S$ with
$$S = S[g_{ij}] + S[\rho] + S_\g.$$
We have $\nabla S_\g = 0$ trivially and $\nabla S[g_{ij}] = \nabla S[\rho] = 0$ since $g_{ij}$ and $\rho$ are given by globally defined tensors on $X$. Thus our global action functional $S[g]$ satisfies
\begin{equation}
\nabla S[g] = 0. \label{eq:Sflat}
\end{equation}
In Section \ref{SecGlobal}, we will express (\ref{eq:Sflat}) as a classical master equation and consider the quantum theory of the nonlinear sigma model globally.

In what follows, we consider only the local theory, i.e., the theory linearized about a fixed $p \in X$. Thus, we will be considering the problem of quantizing the classical $\g$-symmetry of  the  nonlinear sigma model in the case when $X = G/H$.\\

\noindent\textbf{Note:} Henceforth, we take $\Sigma = \R^2$ and all our theories (classical and quantum) will be translation-invariant. As a consequence, we work with observables which belong to the translation-invariant part of (\ref{eq:localO}) and (\ref{eq:globalO}). Thus, when we perform cohomological computations in Sections \ref{Sec:local}--\ref{Sec:global}, all computations will take place on the target $X$, with the dependence on $\Sigma$ dropping out.

\section{Regularization and Renormalization}

\label{SecRR}

As discussed before, in passing from a classical theory to a quantum theory, the notion of a symmetry changes from being expressed as a solution of a classical master equation to a solution of a quantum master equation (\ref{QME-finite}). However, in the infinite dimensionsal setting of quantum field  theory, the measure $d\mu$ and divergence operator occurring in the naive expression for the quantum master equation (\ref{QME-finite}) are a priori ill-defined. Thus, as is customary in quantum theory, it becomes necessary to define a regulated version of all quantities,  including the quantum master equation itself, and then a renormalization procedure must be implemented to render all quantities finite (at every order in perturbation theory) as the regulating parameter is sent to zero.\footnote{Of course, there are other important considerations such as unitarity and causality (in the Lorentzian setting) that one may wish to consider in addition to the requirements we have described. We will not be considering such issues.}

To see what choices of regularization scheme are natural, it is instructive to see how our field theory behaves when placed on a finite lattice $\Lambda$. We are not concerned here with the precise details of how the action functional is to be defined (which will require some choice of discretization of the differentiation operation and boundary conditions at the ends of the lattice) but rather with how a measure on the space of fields is to be defined. With a finite lattice, we have a finite dimensional space of fields, with one copy of $T_pX \oplus T_p^*[-1]X$ for each lattice point $\lambda \in \Lambda$. The inner product on $T_pX$ arising from the metric on $X$ means that we have a corresponding Lebesgue measure $d\mu_\Lambda$ on the total space of bosonic fields $W_\Lambda = \oplus_{\lambda \in \Lambda} T_pX$. Let $y^{i,\lambda}$ be orthonormal coordinates on the copy of $T_pX$ at $\lambda$. It is then possible to define the divergence of a vector field $V = V^{i,\lambda}\pd_{y^{i,\lambda}}$ on $W_\Lambda$ with respect to the measure $d\mu_\Lambda$:
$$\div_{d\mu_\Lambda}(V^{i,\lambda}\pd_{y^{i,\lambda}}) = \sum_{\lambda\in\Lambda}\sum_i\pd_{y^{i,\lambda}} V^{i,\lambda}.$$
If $V$ is translation invariant, i.e., $V^{i,\lambda} = V^i$ is independent of the lattice site $\lambda$, then the above simply becomes the translation-invariant function
$$\div_{d\mu_\Lambda} (V^{i,\lambda}\pd_{y^{i,\lambda}}) = |\Lambda|\pd_{y^i}V^i.$$
From this, one sees that the divergence operator becomes ill-defined as the number of lattice points tends to infinity. Namely, the divergence operator is the contraction of the vector field $V$ with the identity tensor of $W_\Lambda$, and for $\Lambda$ infinite, such a tensor no longer has a well-defined trace.

A regulated divergence operator then has the form of a regulated identity operator. More precisely, in passing to the continuum theory, the integral kernel of the identity operator is formally a delta function, and a regulated version replaces this delta function with a smooth integral kernel that approximates the delta distribution. Such an integral kernel then has a well-defined restriction to the diagonal which can then be integrated (when $\Sigma$ is compact) to yield a well-defined trace.

Our choice of regularization scheme will be the heat kernel method. Namely,
the heat kernel $e^{-t\Delta}$, $t > 0$, with $\Delta$ a Laplace type operator, is taken as a regulated identity operator. In our situation, we take $\Delta$ to be the kinetic term of our field theory, namely the scalar Laplacian $\Delta = -\pd_{x^\mu}\pd_{x^\mu}$ acting on functions on $\Sigma$ with values in $T_pX$. (We will eventually choose normal coordinates at every point, in which case the kinetic operator $S_{kin}$ is of the standard form with $g_{ij,0} = \delta_{ij}$). The integral kernel of $e^{-t\Delta}$ is then given on $\Sigma = \R^2$ by
$$K_t(x,x') = \mr{id}_{T_pX} \otimes \frac{1}{4\pi t}e^{-|x-x'|^2/4t}d^2x',$$
i.e. it is valued in\footnote{When tensoring spaces of smooth functions, we always do so in the following (completed) sense:
$$C^\infty(X) \otimes C^\infty(Y) = C^\infty(X \times Y)$$
for $X$ and $Y$ smooth manifolds. This generalizes to smooth sections of bundles in the obvious way, namely $\Gamma(E_1) \otimes \Gamma(E_2) = \Gamma(E_1 \boxtimes E_2)$. Such tensor products can be reformulated in the setting of nuclear spaces, see \cite{Cos}. The analagous statements carry over if we consider configurations that are distributional instead of smooth.} $C^\infty(\Sigma; T_p X) \otimes C^\infty(\Sigma; T_pX^* \otimes \mr{Dens}(\Sigma))$. 

Picking a length scale $L$, define $K_L$ to be the element of $\Sym^2(C^\infty(\Sigma; T_p X \oplus T_pX^*[-1]))$ given by
$$K_L = \left(\check{e}_i\otimes e_i + e_i \otimes \check{e}_i\right) \otimes \frac{1}{4\pi t}e^{-|x-x'|^2/4t},$$
where $e_i$ and $\check{e}_i$ are orthonormal bases of $T_pX$ and $T_p^*X[-1]$ dual to one another. That is, $K_L$ is obtained from the heat kernel by dropping the translation invariant density factor, shifting the degree of the $T_p^*X$ factor so that $K_L$ has both field and anti-field components, and then symmetrizing the field and anti-field components. This is so that $K_L$, when contracted with local action functionals in the sense of (\ref{eq:contractK}), evaluates in the desired manner on those functionals obtained from multivector fields on $X$ (and such functionals have odd components due to the fact that vector fields on $X$ are graded by their multivector field degree). Define 
\begin{equation}
  \mr{div}_L = \pd_{K_L} + \div_\g. \label{div2terms}
\end{equation}
The first term $\pd_{K_L}$ yields for us our regulated divergence operator in terms of the heat kernel, as parametrized by the length scale $L$. The second term $\div_\g$ is defined to be contraction with the identity tensor in $\g[1] \otimes \g^*[-2]$ and as such is purely finite dimensional and algebraic in nature (it can also be interpreted as a divergence operator on an odd space). It will arise later when we write down the quantum master equation.

Given a vector field $V$ on $X$ expressed in normal coordinates $y^i$ near $p$, $V=V^i\pd_{y^i}$, applying $\mr{div}_L$ to the action functional corresponding to $V$ yields
\begin{align*}
  \div_L S_p[V] %&= \int d^2x \frac{1}{4\pi L}\pd_{y^i}V^i_I(p)\pi^I(x) \\
&= \frac{1}{4\pi L} S_p[\pd_y^i V^i]
\end{align*}
which up to the factor $\frac{1}{4\pi L}$ is the action functional corresponding to the divergence of $V^i\pd_{y^i}$ with respect to the Lebesgue volume form induced on $T_pX$ from the Riemannian metric on $X$. The factor of $\frac{1}{4\pi L}$ in the above is the analog of the factor of $|\Lambda|$ of the lattice regulated divergence operator above. Both these factors blow up in the continuum limit where the length scale or lattice separation tends to zero.

Defining a quantum field theory also requires regulating the propagator for the theory, so that Feynman diagrams are rendered finite. For the kinetic operator $\Delta$ acting on $\pi^i$, the corresponding propagator $P(x,x') = -\mr{id}_{T_pX}\otimes \frac{1}{2\pi}\log |x - x'|$ satisfies
$$\Delta_x  P(x,x') = -\id_{T_pX} \otimes \delta^{(2)}(x-x')$$
in the sense of distributions. Rewriting the propagator as a symmetric field-valued two-tensor, the naive propagator for our theory is  
$$P(x,x') = -\frac{1}{2\pi}\log |x - x'| (e_i \otimes e_i) \in \mathcal{D}(\Sigma; T_pX)^{\otimes 2},$$
where $\mathcal{D}(\Sigma; T_pX)$ denotes the space of distributions on $\Sigma$ valued in $T_pX$. (We express $P$ in terms of the fields because we wish to contract $P$ with local action functionals in Feynman diagrams, and local action functionals take fields as their inputs.) Regulating the propagator involves taming both the ultraviolet divergences (removing the singularity of $P(x,x')$ along the diagonal) and infrared divergences (ensuring rapid decay of $P(x,x')$ at infinity) in order to ensure that all associated Feynman diagrams arising from the perturbative expansion of the theory are finite.

Our method of regulating the propagator will also be via the heat kernel method, so that it will be compatible with the regulation of the divergence operator (in the sense described below). In this method, there are two parameters $\eps, L$ with $0 < \eps < L$, which are ultraviolet and infrared regulating parameters, respectively. The regulated propagator is
\begin{equation}
  P(\eps,L)(x,x') = \int_\eps^L K_t(x,x') e_i\otimes e_i \in C^\infty(\Sigma; T_pX)^{\otimes 2}. \label{prop}
\end{equation}
Since $\Delta P(\eps, L)(x,x') = [K_\eps(x,x') - K_L(x,x')]e_i\otimes e_i$, one sees that that as $\eps \to 0$ and $L \to \infty$, then $P(\eps,L)$ becomes a Green's function for $\Delta$ in the sense of distributions.

The partition function for the nonlinear sigma model expressed in terms of the linearized fields at $p$ (which by abuse of notation we denote also by $\pi$) is formally
$$Z_p = \int D\pi e^{-S_p(\pi)/\hbar}.$$
Thus, the Feynman diagrams of our theory are obtained from the interactions $I = I_p$ appearing in $S_p/\hbar$, as defined in (\ref{interaction}). The propagator placed on the edges of such Feynman diagrams are regulated as above, and hence the regulated Feynman diagrams of our theory are encoded in the functional $e^{\hbar \pd_{P(\eps, L)}}e^{I/\hbar}$. As explained in the appendix, this expression is to be read as the sum of all Feynman diagrams obtained from placing interactions from $I/\hbar$ on the vertices and the propagator $\hbar P(\eps,L)$ on the internal edges. Indeed, $e^{\hbar \pd_{P(\eps, L)}}$ is the operator which implements summing over all possible Wick contractions using the regulated propagator $\hbar P(\eps, L)$. Here we weight the propagator with $\hbar$ so that the weight of connected Feynman diagrams in terms of powers of $\hbar$ coincides with the weight given by the number of loops. Thus, $\hbar$ serves as the perturbative (formal) parameter of the quantum theory.

Since the sum over all Feynman diagrams is equal to the exponential of those which are connected, we have
$$e^{\hbar \pd_{P(\eps, L)}}e^{I/\hbar} = e^{I[\eps,L]/\hbar}$$
where $I[\eps,L]/\hbar$ is a sum over only connected Feynman digrams. It is an action functional valued in formal power series in $\hbar$.\footnote{For $\Sigma = \R^2$, vacuum diagrams are ill-defined, since this involves integrating a constant over all of $\R^2$. Since overall constants are of limited significance, we ignore such diagrams in this case (or more precisely, all statements should be modified to hold modulo constants).} Note that since $I$ contains terms from both the metric and the symmetries arising from $\g$-action, Feynman diagrams are functions of fields and anti-fields and are valued in $C^*(\g)$ (the term $S_\g$, while present, does not participate in Feynman diagrams and goes along for the ride).

Renormalization involves extracting and eliminating the divergences that occur in $I[\eps,L]$ as the ultraviolet regulating parameter $\eps$ is sent to zero.

\begin{Theorem}\cite{Cos}
  There exist local $\eps$-dependent counterterms $I^{CT}(\eps)$ such that $$e^{\hbar\pd_{P(\eps,L)}}e^{(I+I^{CT}(\eps))/\hbar}$$
  has a well-defined limit as a power series functional as $\eps \to 0$.
\end{Theorem}

The Feynman diagrams of the renormalized theory are also obtained from exponentiating only those which are connected.

\begin{Definition}
Define the set of interactions $I[L]$ to be such that
$$e^{I[L]/\hbar} = \lim_{\eps \to 0}e^{\hbar\pd_{P(\eps,L)}}e^{(I+I^{CT}(\eps))/\hbar}.$$
The interactions $I[L]$ are called the \textit{scale $L$ effective interactions}.
\end{Definition}

Thus, it is the set of interactions belonging to $I[L]$ that define the renormalized purely bosonic interactions and renormalized symmetries at scale $L$, well-defined as a formal power series in the perturbative parameter $\hbar$. We only obtain divergent interactions if we try to let the length scale $L$ go to zero.
Passing from scale $L$ to another scale $L'$ involves integrating over quantum fluctuations on length scales between $L$ and $L'$, or in other words, obtaining a set of effective interactions $I[L']$ given by
$$e^{I[L']/\hbar} = e^{\hbar \pd_{P(L,L')}}e^{I[L]/\hbar}.$$
We shall refer to the operation $e^{\hbar \pd_{P(L,L')}}$ as a \textit{change of scale} from scale $L$ to scale $L'$. Observe that this map is reversible, with inverse $e^{-\hbar \pd_{P(L,L')}}$, so that one can change to either higher or lower scales. (Note that while integration is not an invertible process, the operation $e^{-\hbar \pd_{P(L,L')}}$ is invertible since it consists of a sum of the original interactions plus those obtained by Wick contraction.)

\subsection{The Quantum Master Equation}

Given the scale $L$ effective interactions $I[L]$, there is a well-defined notion of a scale $L$ quantum master equation which makes use of the scale $L$ divergence operator $\div_L$. To describe it, we need to introduce some notation. Define
$$Q = -\{S_{kin}, \cdot\}.$$
It is a degree one derivation acting on the space of functionals which arises when one tries to take the Lie derivative of the function $e^{(\pi,Q\pi)/2\hbar}$ appearing in the integrand of the partition function.

For the sake of clarity, we explicitly describe $Q$ acting on local action functionals. From the general framework in the appendix, Poisson bracketing with a local action functional (in particular, the action of $Q$), as an operator acting on the space of local action functionals, factors as a  derivation on $\Sym(\Jet(E)^*)$. In other words, such a Poisson bracket is given by a map $\Jet(E)^* \to \Sym(\Jet(E)^*)$. In our case, $Q$ is given by a $D_\Sigma$-linear map $\Jet(T_pX[1])^* \to \Jet(T_pX)^*$. Let ${\check{e}_i}^* = {\check{e}_i}^*(x)$ and $e_i^* = {\check{e}_i}^*(x)$ denote the constant sections of $\Jet(T_pX)^*$ and $\Jet(T_p^*X)^*$, respectively, dual to the orthonormal basis vectors $e^i$ on $T_pX$ and dual coordinates $\check{e}^i$ on $T_p^*X$, respectively (i.e. $e_i^*(x)(e^j) = \check{e}_i^*(x)\left(\check{e}^j\right) = \delta_{ij}$ for all $x \in \Sigma$). Then we have
\begin{equation}
  Q {\check{e}_i}^* = \pd_\mu^2 e_i^*. \label{eqQ}
\end{equation}
Thus, for example if $V = V^i\pd_{y^i}$ is a vector field, then
\begin{equation}
  (QS_p[V])(\pi) = \int_\Sigma d^2x V^i_I(p)\pi^I\pd_\mu^2 \pi^i. \label{eqQ2}
\end{equation}

The adjoint action of (\ref{eqQ}) yields an action of $Q$ on the space of fields $\cE$. Namely, $Q$ is a degree one map, taking fields to antifields:
\begin{equation}
  Q(f(x)e^i) = (\pd_\mu\pd_\mu f)(x)\check{e}^i. \label{eqQ3}
\end{equation}

\begin{Definition}
  The effective interactions $I[L]$ satisfies the \textit{scale $L$ quantum master equation} if
  \begin{equation}
    (Q + \hbar\div_L)e^{I[L]/\hbar} = 0, \label{QME1}
  \end{equation}
  or equivalently, if
  \begin{equation}
    QI[L] + \frac{1}{2}\{I[L],I[L]\}_L + \hbar \div_LI[L] = 0. \label{QME2}
  \end{equation}
  Here, $\{\cdot,\cdot\}_L$ is the \textit{scale $L$ Poisson bracket} defined by the failure of $\div_L$ to be a derivation on functionals:
  $$\{F,G\}_L := \div_L(FG) - \div_L(F)G - (-1)^{|F|}F\div_L(G).$$
\end{Definition}

Formally, one can interpret (\ref{QME1}) as follows. We have $I[L] = \sum_{i \geq 0} I^{[i]}[L]$, where $I^{[i]}[L]$ consists of terms of $\g$-degree $i$. As with (\ref{QME-finite}), the expression (\ref{QME1}) has terms of $\g$-degree greater than or equal to one. In ghost degree one, (\ref{QME1}) says that the the non-Gaussian ``measure" $e^{I^0[L]/\hbar}e^{(\pi,Q\pi)}D\pi$ on the space of all linear bosonic fields $\pi: \Sigma \to T_pX$ is invariant with respect to the nonlocal scale $L$ vector fields given by $I^1[L]$. In higher ghost degree, we can regard the equation as imposing consistency relations among the vector fields in $I^1[L]$  (the interactions will contain terms up to the maximum ghost degree $\dim \g$). It is difficult to interpret these consistency relations physically but we will not dwell on this issue and treat these relations as formal algebraic consequences of our quantum master equation\footnote{It is well known that the master equation has an interpretation in terms of $L_\infty$-structures \cite{AKSZ}.}.

Observe that $\div_L$ and the scale $L$ Poisson bracket are well-defined at all scales $L > 0$ since $K_L$ is smooth. When $L = 0$, we recover the usual Poisson bracket of Definition \ref{def:PB}, well-defined if at least one of the arguments is a local functional. Note that because $\div_L$ was defined so as to have the term $\div_\g$, then $\{S_\g,\cdot\}_0$ implements the Chevalley-Eilenberg differential as in Section \ref{SecBV}.

The elegance of the above definition is that the quantum master equation at different scales are compatible with the change of scale operation given by $e^{\hbar\pd_{P(L',L)}}$. That is we have the following:

\begin{Lemma}Pick any two scales $L$ and $L'$. Then the scale $L$ effective interactions $I[L]$ satisfy the scale $L$ quantum master equation if and only if the scale $L'$ interactions $I[L']$ satisfies the scale $L'$ quantum master equation.
\end{Lemma}

\Proof This follows from the commutation relation
\begin{equation}
  e^{\hbar \pd_{P(L,L')}}(Q + \hbar\div_L) = (Q + \hbar\div_{L'})e^{\hbar \pd_{P(L,L')}}, \label{eq:CoS}
\end{equation}
which follows from the following observations. First, it is easy to check that
$$[\pd_{P(L,L')},Q] = \pd_{QP(L,L')}$$
where $Q$ acts on $P(L,L')$ via (\ref{eqQ2}) and acting as a derivation. Since $QP(L,L') = K_{L'}-K_L$, then
$$[\pd_{P(L,L')},Q] = \pd_{K_{L'}} - \pd_{K_L}.$$
Since the contraction operators $\pd_{K_L}$ and $\pd_{P(L,L')}$ commute,
$$[e^{\hbar \pd_{P(L,L')}},Q] = \hbar\pd_{K_{L'}}e^{\hbar \pd_{P(L,L')}} - e^{\hbar \pd_{P(L,L')}}\hbar \pd_{K_L},$$
from which (\ref{eq:CoS}) follows.\End

If we consider the quantum master equation (\ref{QME2}) modulo $\hbar$, then we can consider the $L \to 0$ limit from which we recover the classical master equation
$$QI + \frac{1}{2}\{I,I\} = \frac{1}{2}\{S,S\} = 0.$$
However, at order $\hbar$ and higher, the quantum master equation may fail to hold. Indeed, our regularization scheme and introduction of counterterms may violate the equality in (\ref{QME2}), which is to be interpreted as a violation of a classical symmetry at the quantum level.  Thus, define $O[L]$ to be the interaction which measures the obstruction, to leading order in $\hbar$, for $I[L]$ to satisfy the quantum master equation:
\begin{equation}
  \hbar^n O[L] = QI[L] + \frac{1}{2}\{I[L],I[L]\}_L + \hbar \div_L I[L] \mod \hbar^{n+1}. \label{defO}
\end{equation}
In other words,
$$\hbar^n O[L]e^{I[L]/\hbar} = (Q + \mr{div}_L)e^{I[L]/\hbar} \mod \hbar^{n+1}e^{I[L]/\hbar}.$$

\begin{Theorem}\label{ThmO}\cite{Cos}
We have the following
  \begin{enumerate}
    \item $O = \lim_{L \to 0}O[L]$ exists as a functional;
    \item $O$ is a local functional of degree one;
    \item $O$ is $\{S,\cdot\}$ is closed.
  \end{enumerate}
\end{Theorem}

\Proof (Sketch) The first property is true because $O[L']$ is obtained from $O[L]$ by the addition of tree diagrams (any additional loops would lead to a diagram of higher order in $\hbar$, which we mod out by definition). The second property requires some work to prove, but it mainly follows from the fact that the heat kernel $e^{-t\Delta}$ becomes more local (i.e. more concentrated along the diagonal) as $t \to 0$. Thus, the same is true for the propagator $P(\eps,L)$ for $\eps$ and $L$ small. Thus, given the existence of the limiting interaction $O$ by (i), it follows that the limit $O$ must be local (see \cite{Cos} for further details). Finally (iii) follows from
\begin{equation}
  (Q + \hbar \div_L)^2 = 0. \label{BV^2=0}
\end{equation}
Indeed, we apply $(Q + \hbar\div_L)^2$ to $e^{I[L]/\hbar}$ to deduce that $O[L]$ satisfies
$$QO[L] + \{I^0[L],O[L]\}_L = 0,$$
where $I^0[L]$ denotes the order $\hbar^0$ part of $I[L]$ (i.e. only tree diagrams). Now let $L \to 0$ to deduce that
$$QO + \{I,O\} = \{S,O\} = 0.$$
It remains to prove (\ref{BV^2=0}). First, $Q^2 = 0$ since $\{S^{kin}, S^{kin}\} = 0$, since $S$ is purely a function of the fields. We have $\div_L^2 = 0$ since $\pd_{K_L}^2 = 0$ by skew-symmetry ($K_L$ is an odd tensor). Finally, $[Q,\div_L] = \pd_{QK_L} = 0$, since $QK_L=0$ ($Q$ commutes with $e^{-L\Delta}$).

If $O$ is $\{S,\cdot\}$ exact, then it is possible to modify the counterterms $I^{CT}(\eps)$ of order $\hbar^n$ defining $I[L]$ so as to remove the obstruction $O$, i.e., so that the new set of effective interactions $\tilde I[L]$ solves the quantum master equation to order $\hbar^n$. Moreover $\tilde I[L] = I[L] \mod \hbar^n$. The set of all such $\tilde I[L]$ is a torsor for the space of degree zero local action functionals that are $\{S,\cdot\}$ closed since the counterterms defining $\tilde I[L]$ can be modified by precisely such local action functionals. However, if we can find $\tilde I[L]$ which solves the quantum master equation to order $\hbar^n$, i.e., we can eliminate the obstruction, our new set of interactions may fail to solve the quantum master equation to order $\hbar^{n+1}$. We then get a new obstruction which is $\{S,\cdot\}$ closed as before.

In this way, we see that the problem of renormalization while maintining the $\g$-symmetry of the classical theory, i.e., of solving the quantum master equation, becomes a cohomological problem order by order in $\hbar$. At each order, we can solve the quantum master equation precisely when the obstruction to solving it is trivial cohomologically. We record the above remarks in the following:

\begin{Lemma}
  Consider the chain complex of local action functionals $\cO_{\loc,p} = (\cO_\loc(\cE_p), \{S_p,\cdot\})$ for the nonlinear sigma model of maps into an infinitesimal neighborhood of $p \in X$. The space of potential obstructions to the quantum master equation is given by $H^1(\cO_{\loc,p})$. The deformation space of equivalence classes of quantizations, order by order in $\hbar$, is $H^0(\cO_{\loc,p})$.
\end{Lemma}

Note that in the last line above, the deformation is $H^0(\cO_{\loc,p})$ and not closed functionals of degree zero, since we regard exact deformations as being given by the action of an infinitesimal canonical transformation. In our present situation, the image of $\{S_p,\cdot\}$ in degree zero is given by taking a vector field and taking the Lie derivative with $S_p = S_p[g_{ij}] + S_p[\rho]$. Quotienting out by such exact functionals amounts to ignoring the effect of infinitesimal diffeomorphisms on the metric tensor $g_{ij}$ and vector fields determined by $\g$.

\begin{Remark}\label{RemG}
  From a certain point of view, our quantum master equation is a bit naive. Indeed, if we return to the lattice regularization scheme, what we have done is consider the measure $\prod_{\lambda \in \Lambda} d\pi(\lambda)$ on the space of fields $\pi: \Lambda \to T_pX$, where $d\pi(\lambda)$ is the Lebesgue measure on the copy of $T_pX$ at $\lambda$. Thus, we have already destroyed the $\g$-invariance from the outset, since the natural $\g$-invariant measure to use is $\prod_{\lambda \in \Lambda} d\mu_G(\pi(\lambda))$ where $d\mu_G$ is the $G$-invariant volume form on $X$ (formed out of the $G$-invariant metric) pulled back to $T_pX$ via the coordinate system at $p \in X$. (Since we are working in perturbation theory, we really mean the Taylor series of such a volume form at the origin of $T_pX$). Let $\prod_{\lambda \in \Lambda} d\mu_G(\pi(\lambda)) = \prod_{\lambda \in \Lambda} J(\pi(\lambda))d\pi(\lambda)$, so that $J(\pi)$ is the Jacobian factor at each lattice site $\lambda$. A $\g$-invariant quantum measure would be
  $$\prod_{\lambda \in \Lambda} d\mu_G(\pi(\lambda))e^{S[g](\pi(\lambda))/\hbar} = \prod_{\lambda \in \Lambda} d\pi(\lambda) e^{\left(S[g](\pi) + \hbar \sum_{\lambda \in \Lambda} \log J(\pi(\lambda)\right)/\hbar}.$$
  In other words, the Jacobian factor can be absorbed into the interactions as a one-loop effect (there is an $\hbar$ in front of the Jacobian factor). Moreover, since $S[g](\pi)$ is supposed to approximate a local action functional, $S[g](\pi)$ is a sum that is weighted by the square of lattice spacing $\Delta x$ (the volume form $d^2x$ in the continuum theory is discretized to $(\Delta x)^2$ in the lattice theory), whereas the sum $\hbar \sum_{\lambda \in \Lambda} \log J(\pi(\lambda))$ is not weighted by $(\Delta x)^2$. Thus in the continuum limit $\Delta x \to 0$, the Jacobian factor in the exponent contributes a divergent term that is formally proportional to $\delta^{(2)}(0)$ times a local action functional.

What this amounts to is the following. When we work with the naive Lebesgue measure as we did above, we find that a counterterm is needed at one loop to preserve $\g$-invariance of the measure as the ultraviolet regularizing parameter $\eps$ tends to zero. This is the counterterm that would have already been present if we had preserved the Jacobian factor arising from using a $\g$-invariant measure from the outset. This will be made explicit in our analysis of the $O(N)$-model.
\end{Remark}

\section{Local Cohomological Analysis}\label{Sec:local}

In this section, we compute the cohomology of the relevant complex of local action functionals which captures the obstruction and deformation theoretic information to solving the quantum master equation. Here, we focus on the case when $X = G/H$ and $p$ is fixed (the local theory). We return to the global theory where all points of $X$ are considered in Section \ref{SecGlobal}.

Somewhat surprisingly, we find a simple description of the relevant cohomology groups in terms of the Lie algebra cohomology of $\h=\h_p$ with coefficients in a finite dimensional module, where $\h_p$ denotes the Lie algebra of the isotropy subgroup $H_p \subset G$ of $p$. This follows from the following main lemma. Let $\T$ denote some tensor bundle over $X$. Observe that since $\g$ acts on sections of $\T$ by Lie differentiation, it induces an action on $\Jet_p(\T)$. Thus, we may consider the associated Chevalley-Eilenberg cochain complex $C^*(\g, \Jet_p(\T))$. On the other hand, since $H_p$ acts on $\T_p$, the fiber of $\T$ at $p$, we also have an action of $\h_p$ on $\T_p$ and a corresponding Chevalley-Eilenberg cochain complex $C^*(\h, \T_p)$.

\begin{Lemma}\label{LemmaCoho}
  We have a natural chain map
  $$C^*(\g, \Jet_p(\T)) \to C^*(\h, \T_p)$$
  induced by the restriction of $\g$ to $\h$ and the projection $\Jet_p(\T) \to \T_p$. Furthermore, this map is a quasi-isomorphism.
\end{Lemma}

\Proof Pick any coordinate system $y^i$ near $p$ (not necessarily normal coordinates), so that we have an induced algebra isomorphism $\Jet_p(\T) \cong \Sym(T^*_pX) \otimes_\R \T_p$. Choose a complement $\h^\bot$ to $\h$ and a basis $Y_i$ of $\h^\bot$ such that $\rho_*(Y_i)|_p$ is the $i$th coordinate direction $\pd_{y^i}$.

Consider the following decreasing filtration on $C^*(\g, \Jet_p(\T))$. Let $F^k\Jet_p(\T)$ denote those jets which vanish to order $k$ at $p$. Define
$$F^kC^*(\g,\Jet_p(\T)) = \oplus_{k_1+k_2 \geq k}C^{k_1}(\g) \otimes_\R F^{k_2}\Jet_p(\T)$$
Consider the spectral sequence associated to this filtration. The $E^0$ page consists of the vector spaces $F^k/F^{k+1}$ and the differential $d_0$ on $F^k/F^{k+1}$ is essentially the deRham differential:
$$d_0 = \sum Y_i^* \wedge \pd_{y^i}.$$
Indeed, the $\fg$-action on tensors is given by mapping a vector $Z \in \fg$ and tensor $T$ to the Lie derivative $\L_{\rho(Z)}T$. Vectors in $\fh$ vanish at $p$ and are filtration increasing while those from $\fh^\bot$ yield coordinate tangent vectors when evaluated at $p$. Hence, the leading order term in the Lie derivative, and hence the associated Chevalley-Eilenberg differential, is indeed the deRham differential.

The chain complex $(F^k/F^{k+1}, d_0)$ splits as a direct sum of complexes $(C^{k,k_3},d_0)$, $0 \leq k_3 \leq k$, indexed by polynomial degree in $\fh$:
\begin{align*}
  F^k/F^{k+1} &= \bigoplus_{0 \leq k_3 \leq k} C^{k,k_3}\\
  & := \bigoplus_{0 \leq k_3 \leq k} \bigg(\bigoplus_{k_1 + k_2 = k - k_3} C^{k_1}(\fh^\bot) \otimes_\R \left(F^{k_2}\Jet_p(\T)/F^{k_2+1}\Jet_p(\T)\right)\bigg) \otimes_\R C^{k_3}(\fh).
\end{align*}
One can construct a homotopy operator on $F^k/F^{k+1}$
\begin{equation}
  d_0^* = \sum_i y_i Y_i \llcorner \label{eq:d0}
\end{equation}
that preserves each $C^{k,k_3}$ and which satisfies
$$(d_0d_0^* + d_0^*d_0)|_{C^{k,k_3}} = (k - k_3)\mr{id}$$
on $C^{k,k_3}$. Hence for $k_3 < k$, we have that $(C^{k,k_3},d_0)$ is acyclic. Since $d_0$ is zero on $C^{k,k} = C^{k}(\h, \T_p)$, it follows that in taking $d_0$ cohomology and passing to the $E^1$ page, all groups vanish except for $C^*(\h,\T_p)$. Moreover, it is easy to see that the differential $d_1$ is precisely equal to the Chevalley-Eilenberg differential on $C^*(\h, \T_p)$. Indeed, Lie differentiation of a tensor with respect to a vector $Z \in \h$ has two parts: terms which differentiate the tensor and those which differentiate $Z$. The former acts trivially on $\T_p = \Jet^0(\T_p)$ while the latter coincides with the linear action of $\h$ on $\T_p$ induced by the action of isotropy group $H_p$.

What this shows is that the projection $r: C^*(\g, \Jet_p(\T)) \to C^*(\h, \T_p)$ is a chain map and furthermore, it is a quasi-isomorphism. Indeed, the kernel of $r$ is acylic by the above analysis (here we use that $C^*(\g, \Jet_p(\T))$ is complete and Hausdorff with respect to its filtration, so that the collapse of the $E_0$ page implies convergence \cite{Wei}). Thus, an element $c \in C^*(\g, \Jet_p(\T))$ is exact if and only if $r(c)$ is exact, where if $r(c)$ has a primitive $b$, then any element of $r^{-1}(b)$ is cohomologous to $c$.\End

By scaling arguments, one can deduce that the potential obstruction $O$ to the quantum master equation must lie in $C^1(\g,\Jet_p(\Sym^2(T^*X))) \oplus C^2(\g, \Jet_p(T[1]X))$. Indeed, our classical action functional $S$ is invariant under the rescaling of the fields
\begin{equation}
\begin{aligned}
  \pi^i(x) & \mapsto \pi^i(\ell x) \\
  \check{\pi}^i(x) & \mapsto \ell^2 \pi^i(\ell x), \qquad \ell > 0.
\end{aligned} \label{eq:scaling}
\end{equation}
It follows from \cite{Cos} that in quantizing our theory, we can always choose counterterms $I^{CT}(\eps)$ such that $O$ is also scale-invariant. Furthermore, we can assume invariance of the counterterms, and hence of our effective interactions, under the group
$$\mr{Iso}(\R^2) = O(2) \ltimes\R^2$$
of Euclidean isometries. Indeed, both $S_{\mr{kin}}$ and the interactions $I$ are separately invariant under $\mr{Iso}(\R^2)$, and so is the regulated propagator $P(\eps,L)$. It follows that $O$, being $\mr{Iso}(\R^2)$-invariant, is either linear in the anti-field, in which case it is determined by an element of $\Jet_p(T[1]X)$, or else it is a function of the purely bosonic fields and contains two spatial derivatives in which case it is determined by an element of $\Jet_p(\Sym^2(T^*X))$. Since $O$ must have total degree one, the ghost degree of any term of $O$ is determined by its tensor component. This establishes the form of $O$.

Similarly, we can demand that the deformation space of our quantization consists only of  $\mr{Iso}(\R^2)$-invariant, degree zero local action functionals. Additionally, requiring our effective interactions to be \textit{renormalizable} \label{renorm} also requires that we restrict to scale-invariant\footnote{One could also consider terms of positive scaling dimension, given by action functionals that are purely polynomial in the field or consist of one derivative. The latter yield terms which do not respect rotational symmetry and can be ignored. The former can also be ignored since when considering $G$-invariance, functions which are $G$-invariant on $X$ must be identically constant. Hence it suffices to consider only those action functionals we have described.} deformations (which also ensures that the potential obstruction to solving the quantum master equation to the next order in perturbation theory remains scale-invariant). Altogether, our deformation space to quantization, order by order, is $C^0(\g,(\Jet_p\Sym^2(T^*X))) \oplus C^1(\g, \Jet_p(T[1]X))$.\\

\noindent\textit{From now on, we implicitly assume scale invariance and $\mr{Iso}(\R^2)$ invariance of our quantization.}\\

Up to this point, we have not placed any significant constraints on our choice of coordinates at $p$. It is here where we impose the choice of normal coordinates. The significance of this is that the exponential map at $p$ intertwines the $H_p$ action on $X$ with the linear $H_p$ action on $T_pX$ (rotating the tangent vector of a geodesic starting at $p$ rotates the geodesic in the corresponding manner on $X$). What this means is that in normal coordinates, the vector fields $\h_p$ are given by linear vector fields, i.e., they are elements of $T_p^*X \otimes T_pX$. Moreover, $S$ is invariant under the natural $H_p$ action given by the adjoint action on $\g$ and the linear $H_p$ action on $T_pX$ and hence $T_p^*X$. In other words, a rotation of all fields, ghosts, and their anti-fields by an element of $H_p$ preserves $S$.

Consequently, we can quantize our theory while imposing $H_p$-invariance on the effective interactions:

\begin{Lemma}\label{LemmaHinv}
  We can always choose the counterterms $I^{CT}(\eps)$ defining the set of effective interactions $I[L]$ to be $H_p$-invariant and vanish when evaluated with an element of $\fh_p$. Consequently, we can always choose $I[L]$ to be invariant under the natural action of $H_p$.
\end{Lemma}

\Proof Since $H_p$ acts linearly on fields, it preserves the kinetic and interaction parts of the bosonic action separately. Thus, $H_p$ preserves the propagator $P(\eps,L)$ and so commutes with the operation of Wick contraction $e^{\hbar\pd_{P(\eps,L)}}$. It follows that counterterms $I^{CT}(\eps)$ can be taken to be $H_p$-invariant as well simply by averaging over $H_p$ if necessary.

Finally, since $S|_{\h_p}$ is linear in the bosonic fields, no counterterms are needed to renormalize diagrams involving vertices from $S|_{\h_p}$, since diagrams containing external trees are already rendered finite by previous counterterms that renormalize subdiagrams.\End

\begin{Definition}\label{DefHinv}
  We say that a set of effective interactions $I[L]$ is \textit{strongly $H_p$-invariant} if the counterterms which define it satisfy the properties within Lemma \ref{LemmaHinv}. Equivalently, $I[L]$ is strongly $H_p$-invariant if $I[L]$ is invariant under the natural action of $H_p$ and such that the $\h$-dependence of $I[L]$ enters only through the attachment of tree diagrams arising using vertices from $S[\rho]|_\fh$.
\end{Definition}

We include the latter definition in the above to show that it is possible to define what it means for a set of effective interactions to be strongly $H_p$-invariant without reference to counterterms. (Indeed, one perspective of effective field theories is that no reference should be made to counterterms as the latter are not physically observable.)

\begin{Theorem}\label{ThmSec4}
  Let $I[L]$ be a set of strongly $H_p$-invariant effective interactions. Then the solution to the quantum master equation order by order in $\hbar$ is unobstructed. Given a solution $I[L]$ to the quantum master equation modulo $\hbar^n$, the space of strongly $H_p$-invariant solutions $\tilde I[L]$ such that $\tilde I[L] = I[L] \mod \hbar^{n-1}$ is a torsor with respect to the vector space $\mr{Met}^G$.
\end{Theorem}

\Proof Via scale-invariance and $\mr{Iso}(\R^2)$-invariance, the relevant obstruction-deformation complex is
\begin{equation}
  \begin{CD}
  \Jet_p(T[1]X) @>>> \Jet_p(\Sym^2(T^*X)) \\
  @VVV  @VVV \\
  C^1(\g, \Jet_p(T[1]X)) @>>> C^1(\g,\Jet_p(\Sym^2(T^*X))) \\
  @VVV  @VVV \\
  C^2(\g, \Jet_p(T[1]X)) @>>> C^2(\g,\Jet_p(\Sym^2(T^*X))) \\
@VVV  @VVV \\
\vdots  & & \vdots \\ \\
\end{CD}\label{eq:Jetcomplex}
\end{equation}
The differential $\{S,\cdot\}$ splits into a vertical differential $\{S_\rho + S_\g,\cdot\}$ which yields the Chevalley-Eilenberg differential of the columns and a horizontal differential $\{S_0,\cdots\}$ which computes the Lie derivative of the classical metric with respect to a multivector field.

When we compute vertical cohomology first, by Lemma \ref{LemmaCoho}, we are left with the complex $H^*(\h, T_pX[1]) \overset{\{S_0,\cdot\}}{\longrightarrow} H^*(\h, \Sym^2(T_p^*X))$, where $\{S_0,\cdot\}$ is the induced differential on the Lie algebra cohomology groups. We are interested in the cohomology groups in degree zero and one. For degree zero, this is given by
\begin{equation}
  \coker \Big(H^0(\h, T_pX[1])\to H^0(\h, \Sym^2(T_p^*X))\Big) \oplus \ker \Big(H^1(\h, T_pX[1]) \to H^1(\h, \Sym^2(T_p^*X))\Big)
\end{equation}
We can ignore the second space, since strongly $H_p$-invariant quantizations have no quantum corrections to the linear $\h$ vector fields. For the first space, observe there is a one-to-one correspondence between $H_p$-invariant elements of $\T_p$ and $G$-invariant sections of $\T$ (for any tensor bundle on $\T$ on $X$). Thus, the above cokernel is just the space of $G$-invariant elements of $\Sym^2(T^*X)$ on $X$ modulo Lie derivatives of the classical metric $g_{ij}$ by $G$-invariant vector fields, i.e. $\mr{Met}^G$.

For cohomology in degree one, there is in general an obstruction space since $H^1(\h, \Sym^2(T_pX))$ can be nonzero (for instance if $\h$ is abelian, this space is spanned by the $H_p$ invariant elements of $\Sym^2(T_pX)$) as well as $H^2(\h, T_pX[1])$. However, we will show that given a strongly $H_p$-invariant set of effective interactions $I[L]$, the leading order obstruction $O$ satisfies $O|_\fh=0$. This will show that it is cohomologically trivial by Lemma \ref{LemmaCoho}. Moreover, by averaging with respect to $H_p$, the cocyle which kills $O$ can be chosen $H_p$-invariant and annihilated when evaluated against vectors in $\h_p$ (the latter statement can be seen from the fact that we found a homotopy operator $d_0^*$ (\ref{eq:d0}) which is $H_p$-equivariant).

The definition of $O[L]$ in (\ref{defO}) is very unwieldy, since it involves the $I[L]$, which involve a sum over infinitely many nonlocal interactions. We have the more convenient expression
$$\hbar^n O[L] = \lim_{\eps \to 0}e^{\hbar\pd_{P(\eps,L)}}(Q + \Delta_\eps)e^{(I + I^{CT}(\eps))/\hbar} \mod \hbar^{n+1}e^{I[L]/\hbar}$$
using (\ref{eq:CoS}). Consider $(Q + \Delta_\eps)e^{(I + I^{CT}(\eps))/\hbar}|_\h =: V(\eps)e^{(I + I^{CT}(\eps))/\h}|_\h$. Writing $V(\eps) = V^1(\eps) + V^2(\eps)$ in terms of its components $V^i(\eps)$ of ghost degree $i$, we have
$$V^1(\eps) = \left(QS_\rho + \{S_\rho, I^0 + I^{CT,0}(\eps)\}_\eps + \hbar \div_\eps S_\rho\right)|_\h$$
and
$$V^2(\eps) = \left(\{S_\g, S_\rho\} + \frac{1}{2}\{S_\rho, S_\rho\}_\eps\right)|_\h.$$
For $V^2(\eps)$, this vertex is linear in the bosonic field, so its $\eps \to 0$ limit vanishes and contributes nothing to $O[L]|_\h$ for $L > 0$. For $V^1(\eps)|_\h$, we have $\div_\eps S_\rho|_\h = QS_\rho|_\h = 0$ and we have just the remaining term $\{S_\rho|\h, S^0 + I^{CT,0}(\eps)\}_\eps$. There are two cases for how this vertex contributes to the diagrams appearing in $O[L]$. We have diagrams for which the remaining bosonic edge of $S_\rho|_\h$ in $V^1(\eps)$ participates in a Wick contraction, and those for which it does not. For those that do not, then all such diagrams contribute $\lim_{\eps \to 0}\{S_\rho|_\h, e^{I[\eps,L]/\hbar}\}_\eps$ to $O[L]e^{I[L]/\hbar}$, which vanishes for $\eps = 0$ by $H_p$-invariance of $I[L]$. For the diagrams in the remaining case, we proceed as follows:

Pick any $Z \in \h$. The vector field $\rho(Z)$ is a linear vector field of the form $a_{ij}y^i\pd_{y^j}$ for some matrix $a_{ij}$ (in fact, $a_{ij}$ are skew-symmetric since the vector fields $\pd_{y^i}$ form an orthonormal basis for $T_pX$). Hence, $S_\rho(Z)$ is an action functional of the form $\int d^2x a_{ij}\pi^i\check{\pi}^j$. The anti-field becomes contracted with one leg of $K_\eps$ via the bracket $\{\cdot, \cdot\}_\eps$. If the remaining bosonic field receives a Wick contraction, then the $S_\rho(Z)$ vertex contributes
$$\int d^2x a_{ij}P(\eps,L)(x,y)K_\eps(x,z)\pd_{\pi^i(y)}\pd_{\pi^j(z)}$$
to Feynman diagrams, where $\pd_{\pi^i(y)}$ denotes the placement of the free leg of the propagator $P(\eps,L)(x,y)$ at an edge $\pi^i(y)$ of some other local action functional whose integration variable is $y$ and similarly for $\pd_{\pi^j(z)}$. But observe that
\begin{align*}
  \int d^2x a_{ij}P(\eps,L)(x,y)K_\eps(x,z)\pd_{\pi^i(y)}\pd_{\pi^j(z)} =
  \hspace{2in} \\ \int d^2x P(\eps,L)(x,y)\int d^2w a_{ij}K_\eps(x,w)K_0(w,z)\pd_{\pi^i(y)}\pd_{\pi^j(z)}
\end{align*}
since $K_0(w,z) = \delta^{(2)}(w - z)$. But the above expression becomes
$$\int d^2x P(\eps,L)(x,y)K_\eps(x,w)\pd_{\pi^i(y)} \pd_{\pi^i(w)}\left(\int d^2w a_{ij}\pi^i(w) K_0(w,z)\pd_{\pi^j(z)}\right)$$
and the expression inside the parenthesis implements the action of $Z$ on fields. By $H_p$-invariance, this operator vanishes and hence the corresponding diagrams vanish.
\End

\section{Example: The $O(N)$-model}

\label{SecON}

We perform some explicit computations for the case $X = S^{N-1} = O(N)/O(N-1)$. This allows us to supplement the abstract cohomological considerations above with a concrete example. Furthermore, we are able to relate our work to the old work of \cite{BLZ} which studied the perturbative renormalization of the rotational symmetry of the $O(N)$-model. In a sense, our work is a generalization and rigorization of their analysis. Indeed, in \cite{BLZ} it is assumed that there is no anomaly for the $O(N)$ symmetry and counterterms for the theory are chosen, inductively order by order in perturbation theory, so as to maintain this symmetry. In our work, counterterms are chosen first and then adjusted so that the quantum master equation is satisfied. This latter approach is more satisfactory since it separates the distinct procedures of renormalization and removing potential anomalies. The former is always possible, but the latter may be cohomologically obstructed a priori. Nonetheless, in agreement with \cite{BLZ}, we find that there is no anomaly for the $O(N)$ symmetry, which for us is a special case of our more general result for homogeneous spaces.

Our goal here is to understand explicitly the one-loop counterterms that occur in the renormalization process of the $O(N)$ model. This is because in much of the literature on renormalization, a symmetry of a quantum theory is usually expressed in terms of a corresponding symmetry of the counterterms needed for the theory. In some sense, however, this way of proceeding can be unsatisfactory for several reasons. First, counterterms are regularization scheme dependent and not physically observable quantities. Thus, it is awkward to express a symmetry in terms of quantities that diverge instead of those which are observable. Second, it is possible for a quantum field theory not to require any counterterms but which still has an anomaly \cite{AS, CosWG}.

In our present approach, by working with a (regulated) quantum master equation expressed solely in terms of effective interactions, the notion of a symmetry of the quantum theory is expressed solely in terms of well-defined entities that enter into measurable quantities. The downside with the quantum master equation is that it consists of nonlocal effective interactions, which consist of an infinite sum of complicated Feynman diagrams. On the other hand, counterterms, at least at one loop, are usually easier to analyze than the effective interactions. Thus, while counterterms are auxiliary quantities that can be argued (as above) to not play a fundamental role in the notion of symmetry, much insight can be gleaned from how symmetry constrains the form of the counterterms (as is well known).

We carry out such an analysis at one loop for the $O(N)$-model. First, we set up some notation. Although our previous analysis was done with respect to normal coordinates about some point $p$, we could have worked with any coordinate system in which the vector fields arising from $\fh=\fh_p$ become linear vector fields. In the case at hand, let $p = (0,\ldots, 0,1)$ be the north pole of
$$X = S^{N-1} = \{(y^1,\ldots, y^{N-1}, \sigma) \in \R^N: \sum (y^i)^2 + \sigma^2 = 1\}.$$
Consider ``graph coordinates" for $T_pX$, where a point $\vec{y} = (y^1,\ldots, y^{N-1})$ is mapped to the corresponding point $(\vec{y}, \sigma)$ on $X$, where $\sigma = \sigma(y) = \left(1 - |\vec{y}|^2\right)^{1/2}$. We have $\g = \frak{so}(N)$ is spanned by the elements $Z_{ij}$, $i,j = 1,\ldots, N$, $i \neq j$, where $Z_{ij}$ is the generator that rotates the $i$th coordinate direction into the $j$th. That is,
$$Z_{ij} = y^i\pd_{y^j} - y^j\pd_{y^i}.$$
(To simplify notation, in what follows, we will not always distinguish carefully between elements of $\g$ as ``external" elements of $\frak{so}(n)$ and as vector fields on $S^{N-1}$. We will also not carefully distinguish between vector fields and their jets at $p$.) The subalgebra $\h_p$ corresponds to the span of $Z_{ij}$, $i, j < N$ and we have the complementary vector fields which form the space $\h_p^\bot$ spanned by the $Z_{Ni}$. When acting on $S^{N-1}$, the vector fields on $Z_{ij}$, pulled back to $T_pX$ via the coordinate system introduced are such that $Z_{ij}$ retain their linear form for $i,j<N$, while the
$$Z_{Ni} = \sigma(y) \pd_{y^i}$$
become nonlinear.

We now consider the nonlinear sigma model of maps from $\Sigma$ into an infinitesimal neighborhood of $p \in S^{N-1}$. We have corresponding fields $\pi^i(x)$, $i = 1,\ldots, N-1$ valued in $T_pX$, and the nonlinear field $\sigma(x) = \sigma(\vec{\pi}(x)) = \sqrt{1 - |\vec{\pi}(x)|^2}$. We can express our action functional $S = S_p$ in terms of the Taylor series expansion of the round metric $g_{ij}$ on $X$ and the vector fields $Z_{ij}$ at the origin of $T_pX$. Explicitly, the round metric is given by
$$\delta_{ij}dy^idy^j + \frac{(y^i dy^i)(y^jdy^j)}{1 - |\vec{y}|^2},$$
which by expanding the denominator can be expressed as a power series of the $y^i$. The above decomposition of the metric into a flat part and the remaining part yields a corresponding decomposition of the bosonic action
\begin{align*}
  S[g_{ij}] &= S_{\mr{kin}}[g_{ij}] - I[g_{ij}] \\
  &= \frac{1}{2}d^2x \int \pd_\mu \pi^i(x)\pd_\mu\pi^i(x) + \frac{1}{2}\int \frac{(\vec{\pi}\cdot\pd_\mu \vec{\pi})(\vec{\pi}\cdot\pd_\mu \vec{\pi})}{1-|\vec{\pi}|^2}
\end{align*}
into its kinetic and interaction terms. Likewise, $S_\rho$ is given by
\begin{align*}
  S_\rho[Z_{ij}] &= \int d^2x\Big(\pi^i(x)\check{\pi}^j(x) - \pi^j(x)\check\pi^i(x)\Big) \\
  S_\rho[Z_{Ni}] &= \int d^2x \sigma(x)\check\pi^i(x).
\end{align*}

Next, we analyze the quantum master equation at one loop. Recall that the scale $L$ differential $Q + \hbar \div_L$ applied to the effective action $e^{I[L]/\hbar}$ can be obtained via
\begin{align}
  (Q + \hbar\div_L)e^{I[L]/\hbar} = \lim_{\eps\to 0}e^{\hbar P(\eps,L)}(Q + \hbar\div_\eps)e^{(I+I^{CT}(\eps))/\hbar},  \label{Oequation}
\end{align}
due to the commutativity property (\ref{eq:CoS}). This is a convenient expression because the right-hand side is expressed in terms of the interactions arising from $(Q + \hbar\div_\eps)e^{(I+I^{CT}(\eps))/\hbar}$, which are essentially local for $\eps$ small. It follows that the one-loop obstruction $O[L]$ is obtained from attaching diagrams to the functional $O_\eps$ defined by
$$O_\eps e^{(I + I^{CT}(\eps))/\hbar} = (Q+\hbar\div_\eps)e^{(I + I^{CT}(\eps))/\hbar} \mod \hbar^2 e^{(I + I^{CT}(\eps))/\hbar},$$
i.e. the functional
$$O_\eps = Q(I+ I^{CT}(\eps)) + \frac{1}{2}\{I+I^{CT}(\eps),I+I^{CT}(\eps)\}_\eps + \hbar \Delta_\eps I \mod \hbar^2.$$
We can write $O_\eps = O_\eps^0 + O_\eps^1$ by collecting terms that are of order $\hbar^0$ and $\hbar^1$, respectively. We consider the contribution to $O[L]$ from the $O_\eps^i$, which from (\ref{Oequation}), is given by
\begin{align}
  O[L]e^{I[L]/\hbar} = \lim_{\eps\to 0}e^{\hbar P(\eps,L)}O_\eps e^{(I+I^{CT}(\eps))/\hbar} \mod \hbar^2 e^{I[L]/\hbar}.
\end{align}
The term $O_\eps^1$ can only receive additional tree attachments, while $O^0_\eps$ can receive one loop corrections (either through self loops or from adding vertices from $I$). Focus on the $O^0_\eps$ terms. Since the classical action satisfies the classical master equation, we have
$$O^0_\eps = \frac{1}{2}\{I,I\}_\eps - \frac{1}{2}\{I,I\}_0 = \left(\{S[\rho], I[g]\}_\eps -  \{S[\rho], I[g]\}_0\right) + \frac{1}{2}\left(\{S[\rho],S[\rho]\}_\eps - \{S[\rho],S[\rho]\}_0\right).$$
For the first term, we have an interaction which is of order $\eps$ and a loop creates a divergence of order at most $\eps^{-1}$ (we have $P(\eps,L)(x,x)$ is of order $\log \eps$ while $\pd_x^2 P(\eps,L)(x,x)$ is of order $\eps^{-1}$). For the second term, we have an interaction which is of order $\eps$ and a loop creates an integral kernel of order at most order $\log \eps$ (since for this second term, none of the interactions have derivatives to create higher order divergences). Altogether, we find that the $\eps \to 0$ limit of the diagrams of $O[L]$ arising from $O^0_\eps$ are finite. It follows that diagrams arising from $O_\eps^1$ have a finite $\eps \to 0$ limit as well. But then $\lim_{\eps \to 0}O^1_\eps$ must be finite, since diagrams arising from $O^1_\eps$ consist of tree attachments, and these never create divergences.

It is the existence of $\lim_{\eps \to 0}O^1_\eps$ that leads to relations among the counterterms arising from the $\g$-symmetry of the classical action. We group the terms of $O^1_\eps$ by the types of local action functionals that are involved. Indeed, the local counterterms are either corrections to the metric, the volume form, or the vector fields arising from the $\g$-action, i.e., they contain either two derivatives of the bosonic field, no derivatives, or are linear in the anti-field. Call these counterterms $I^{CT, \mr{met}}$, $I^{CT,\mr{fun}}$, and $I^{CT,\mr{vec}}$, respectively. In what follows, we write $\equiv$ to mean equality modulo terms that have a finite $\eps$ limit. Thus, we have $O^1_\eps \equiv 0$, and it decomposes into three separate equations:\\

(I) $\{S[\rho], I^{CT, \mr{fun}}(\eps)\}_\eps + \hbar \div_\eps S[\rho] \equiv 0$.\\

We compute both of terms of the left-hand side explicitly and will see that they cancel. Observe that this expresses, in a regulated fashion, the $\g$-invariance of the natural $\g$-invariant measure on the space of fields, as opposed to the naive Lebesgue measure, at one loop (see Remark \ref{RemG}). We only need to verify (I) for the nonlinear symmetries $Z_{Ni}$. We know that $S[\rho](Z_{Ni}) = \int d^2x \sigma(x)\check{\pi}_i(x).$ Hence, $\div_\eps S[\rho](Z_{Ni}) = \frac{1}{4\pi \eps}\int d^2x \pd_{\pi^i}\sigma(x)$. On the other hand, $I^{CT, \mr{fun}}$ is the one-loop diagram obtained by (i) taking $n$ copies of $I[g]$ and contracting the edges that contain derivatives of the bosonic field cyclically to form an $n$-vertex wheel; (ii) summing over all $n$. When two derivatives hit a propagator $P(\eps,L)$, one obtains the difference in heat kernels $K_\eps - K_L$, of which only $K_\eps$ contributes to the $\eps \to 0$ divergence. Since $K_\eps = \delta^{(2)}(x-x') + O(\eps)$ as a distribution, the $n$-vertex diagram we obtain from (i) is modulo terms that have an $\eps \to 0$ limit
$$\frac{1}{2n}\int d^2x F(\pi(x))^n K_\eps(x,x) = \frac{1}{4\pi \eps}\frac{1}{2n}\int d^2x F(\vec\pi(x))^n,$$
where $2n$ is the symmetry factor of a wheel with $n$ vertices and
$$F(\vec\pi(x)) = -\frac{|\vec\pi(x)|^2}{1-|\vec\pi(x)|}$$
is the interaction obtained from $I[g_{ij}]$ by amputating derivative legs.

Summing over all $n$, we obtain
\begin{align*}
  \frac{\hbar}{4\pi\eps}\sum_{n=1}^\infty \frac{F(\vec\pi)^n}{2n} &= \frac{\hbar}{4\pi\eps}
  \left(-\frac{1}{2}\log(1 - F(\pi))\right) \\
  &= \frac{\hbar}{4\pi\eps}\left(-\frac{1}{2}\log\left(\frac{1}{1-|\vec\pi|^2}\right)\right) \\
    &= \frac{\hbar}{4\pi\eps}\log \sigma(x)
\end{align*}
which means
$$I^{CT,fun}(\eps) = -\frac{\hbar}{4\pi\eps}\int d^2x \log \sigma(x)$$
modulo finite terms.

We can replace $\{\cdot,\cdot\}_\eps$ with $\{\cdot,\cdot\}_0$ in (I) since we are only interested in terms up to those with finite $\eps \to 0$ limits. Thus, the first term $\{S[\rho](Z_{Ni}), I^{CT,\mr{fun}}(\eps)\}_0$, which computes the effect of rotation by $Z_{Ni}$ is given by
$$-\frac{\hbar}{4\pi \eps}\int d^2x \sigma \pd_{\pi^i} \log \sigma(x) = -\frac{\hbar}{4\pi \eps}\int d^2x \pd_{\pi^i}\sigma(x).$$
This precisely cancels $\hbar \div_\eps S[\rho](Z_{Ni})$.\\

(II) $\{S[\rho]+S_\g, I^{CT, \mr{vec}}(\eps)\}_\eps \equiv 0$.\\

Replacing $\{\cdot,\cdot\}_\eps$ with $\{\cdot,\cdot\}_0$, this says $I^{CT,\mr{vec}}(\eps)$ is a first order deformation of the Lie algebra homomorphism $\rho$. That is, letting $c^1 = I^{CT,\mr{vec}}: \g \to \Jet_p(TX)$, we get\footnote{We will abuse notation in what follows by not carefully distinguishing between elements of $\g$ and the (jet of) their image under $\rho$.}
\begin{equation}
  Z\cdot c_1(Z') - Z' \cdot c_1(Z) - c_1([Z,Z']) = 0, \qquad Z,Z' \in \g. \label{eq:c1}
\end{equation}
We could determine $c^1$ directly through a Feynman diagrammatic analysis, but in the spirit of our cohomological analysis, we solve the above equations purely Lie algebraically. Of course, any exact $1$-cochain satisfies the condition (\ref{eq:c1}) of being closed, so we will seek only those solutions that satisfy those constraints imposed by our quantization procedure.

Solving for $c_1$ is greatly facilitated by the fact that it obeys many identities. Namely $c_1|\fh = 0$ and
\begin{align*}
  [\h,\h^\bot] & \subseteq \h^\bot \\
  [\h^\bot,\h^\bot] & \subseteq \h.
\end{align*}
The last equality follows from $S^{N-1}$ being a symmetric space. We obtain two types of equations for $c_1$ from (\ref{eq:c1}) corresponding to whether one or both vectors lie $\h^\bot$, respectively:
\begin{align}
  [Z_{ij},c_1(Z_{Nj})] - c_1([Z_{ij},Z_{Nj}]) &= 0 \label{eq:c1-1} \\
  [Z_{Ni},c_1(Z_{Nj})] - [Z_{Nj},c_1(Z_{Ni})] &= 0, \label{eq:c1-2}
\end{align}
for all $i,j < N$. Since $H_p$ acts irreducibly on $\fh^\bot$ and  (\ref{eq:c1-1}) implies $c_1$ is $H_p$-equivariant, $c_1$ is determined by its value along a single direction, say $Z_{N1}$. Since the $Z_{ij}$ and $Z_{N1}$ commute for $1 < i,j < N$, one also deduces from (\ref{eq:c1-1}) that $[Z_{ij}, c_1(Z_{N1})] = 0$. We know that the counterterm for $\rho(Z_{Ni})$ must be a vector in the $y^i$ direction, and hence $c_1(Z_{N1}) = C Z_{N1}$ for some function $C$ on $S^{N-1}$. By $H_p$-invariance, $c_1(Z_{Ni}) = C Z_{Ni}$ for all $i$, with $C$ being $H_p$-invariant.

Plugging this in to (\ref{eq:c1-2}) and using the relation
$$Z_{ij} = \frac{y^i}{\sigma}Z_{Nj} - \frac{y^j}{\sigma}Z_{Ni}$$
which express the vectors from $\h$ in terms of the local tangent frame provided by the $Z_{Ni}$, we obtain the differential equation
\begin{align}
  \pd_{y^i} C &= \frac{2y^i}{\sigma^2}C
\end{align}
which has a unique solution solution up to an overall constant
$$C = \lambda\sigma^{-2}, \qquad \lambda \in \R.$$
Thus, we have determined $c^1(Z_{Ni}) = C Z_{Ni}$ and so the local action functional corresponding to $I^{CT,\mr{vec}}(\eps)$ is proportional to
$$\hbar \int d^2x C(\vec{\pi}(x))\sigma(x)\check{\pi}^i(x) = \hbar \lambda \int d^2x \sigma(x)^{-1}\check{\pi}^i(x).$$
The divergent $\eps$-dependent coefficient for the counterterm is easily seen to be proportional to $\log \eps$.\\

(III) $\{S[\rho], I^{CT, \mr{met}}(\eps)\}_\eps + QI^{CT, \mr{vec}} + \{I^{CT, \mr{vec}}(\eps), I[g_{ij}]\}_\eps \equiv 0$. \\

As before, replacing $\{\cdot,\cdot\}_\eps$ with $\{\cdot,\cdot\}_0$, this equation becomes
$$\{S[\rho], I^{CT, \mr{met}}(\eps)\} + \{S[g_{ij}], I^{CT, \mr{vec}}(\eps)\} = 0.$$
Having solved for $I^{CT,\mr{vec}}$ above, one deduces that $I^{CT, \mr{met}}$ is the functional corresponding to the metric
\begin{equation}
  \mu g_{ij} - 2\lambda\frac{(y^i dy^i)(y^jdy^j)}{\sigma^4}, \label{CTmet}
\end{equation}
where $\lambda$ is the same constant appearing in $I^{CT, \mr{vec}}$ and $\mu$ some other constant. This is most easily seen as follows. We know that the cochain $c^1$ determining $I^{CT, \mr{vec}}$ is closed, hence exact by (\ref{LemmaCoho}). What this means is that there is a vector field $W$ defined near $p$ such that $c^1(Z) = [W, \rho(Z)]$. In other words, the counterterm $I^{CT, \mr{vec}}$ can be removed by an the action of an infinitesimal diffeomorphism $W$ on the fields, i.e. by a field strength renormalization. One readily checks that $W$ is given by the radial vector field $-\lambda y^i\pd_{y^i}$ on $T_pX$. On the other hand, by Theorem \ref{ThmSec4}, the equivalence classes of deformations to our theory are $O(N)$-invariant metrics on $S^{N-1}$, which is just a one-dimensional space spanned by multiples of the round metric. It follows that the metric counterterm $I^{CT, \mr{met}}$, modulo finite terms, is exact modulo terms proportional to the round metric. Thus, $I^{CT, \mr{met}}$ has the form (\ref{CTmet}), since
$$\L_W g_{ij} = -2\lambda \left(g_{ij} + \frac{(y^i dy^i)(y^jdy^j)}{\sigma^4}\right).$$

Our analysis of one-loop counterterms for the $O(N)$-model reproduces the results of \cite{BLZ}. Altogether, our results can be interpreted as saying that to all loop order, there is a single renormalized coupling constant, namely, the round metric, modulo field strength renormalization. The renormalized round metric is valued in power series in $\hbar$ and this expresses the $O(N)$ symmetry of the quantum theory.

\section{Global Quantization}\label{Sec:global}

\label{SecGlobal}

Thus far, we have worked perturbatively about a point $p \in X$ and have shown that it is possible to quantize the nonlinear sigma model for fields mapping into an infinitesimal neighborhood of $p$ while preserving the $\g$-symmetry when we specialize to $X = G/H$. In fact, by showing that the corresponding quantum master equation holds, we also obtain that these $\g$-symmetries obey higher consistency relations as a formal consequence.

In this section, we consider the problem of global quantization. Recall that our global action functional, expressed as a section of a jet bundle, is annihilated by a natural flat connection since the action functional arises from the jet of a globally defined metric and, in the case of $X = G/H$, the jet of globally defined vector fields arising from the $\g$-action. Our first task is to encode the action of $\nabla$ in terms of a classical master equation so that we may obtain the corresponding form of the quantum master equation for the quantized theory. It turns out that some subtleties are involved in the latter process \cite{LiLi} and we supply the details here. Notwithstanding, what we find is that the obstruction to solving the quantum master equation is given by the cohomology in degree one of the differential arising from the classical master equation. From this, we find that there is no obstruction to preserving $\nabla$-flatness of the quantum theory. When $X = G/H$, then for $G$ and $H$ satisfying the hypotheses of the main theorem, the potential obstruction to a global $\g$-symmetry lies in $H^1(X; \mr{Met}^G)$. While no hypothesis on $G$ and $H$ were needed to quantize at individual points $p \in X$, our hypotheses on $G$ and $H$ reflect the fact that we need some global constraints to ensure that the family of quantizations for every $p \in X$ is done consistently. In particular, if $H^1(X) = 0$, which is true for $G$ compact and semi-simple, there is no anomaly for both the $\g$-symmetry and $\nabla$-symmetry.

The natural flat connection $\nabla$ on $\Jet(X)$ induces a flat connection on $\wSym(T^*X)$ via the family of normal coordinates which identities $\Jet(X)$ with $\wSym(T^*X)$.  Since the flat connection on $\Jet(X)$ is a derivation with respect to the natural algebra structure on $\Jet(X)$, the induced flat connection on $\wSym(T^*X)$ is also a derivation with respect to the latter's natural algebraic structure. Thus, $\nabla$ is determined completely by its action on $\Sym^1(T^*X)$. Since $\nabla$, as a connection, satisfies the Leibnitz rule
$$\nabla (fs) = df s + f\nabla s,$$
it follows that if we write
\begin{align*}
\nabla = \sum_{k \geq 0} d_k, \qquad d_k: \Sym^1(T^*X) \to \Omega^1\left(\Sym^k(T^*X)\right),
\end{align*}
then the $d_k$ are $C^\infty(X)$-linear (i.e. are bundle maps) while $d_1$ satisfies the Leibnitz rule (i.e. is a connection on $T^*X$). We record the following (nonessential) fact:

\begin{Lemma}
  If the family of coordinate systems $\Theta$ is such that each $\Theta_p$ is a normal coordinate system at $p$, then $d_1$ is the Levi-Civita connection.
\end{Lemma}

The $d_k$, $k \neq 1$, are bundle maps that together yield a derivation of $\Omega^*\left(\wSym(T^*X)\right)$. Moreover, since $\Sym(T^*X) \otimes TX$ acts on $\Sym(T^*X)$ in the natural way by derivation, i.e., for each $k$, we have the contraction
\begin{align*}
  T_X \otimes \Sym^k(T^*X): &\to \Sym^{k-1}(T^*X) \\
  v \otimes T & \mapsto \pd_v T
\end{align*}
which extends $\Sym(T^*X)$ linearly in the first factor, we can define $d_k: T_X \to \Omega^1\left(\Sym^{k-1}(T^*X)\right) \otimes TX$ by the adjunction formula
$$\pd_T(d_kv) = d_k(\pd_vT) - \pd_v(d_kT), \qquad v \in TX, \quad T \in T^*X.$$
From this, we can extend $d_k$ as a degree one derivation to a map
$$d_k: \wSym(T^*X \oplus T[1]X) \to \Omega^1\left(\wSym(T^*X \oplus T[1]X)\right).$$
It readily follows that there is a functional $I_k \in \Omega^1\left(\Sym^k(T^*X)\right) \otimes T[1]X$ such that
$$-\{I_k,\cdot\} = d_k$$
on $\Sym^*(T^*X \oplus T[1]X)$ (where $\{\cdot,\cdot\}$ is the Schouten-Nijenhuis bracket). Indeed, any derivation $\delta$ on a space of formal functions (such as $d_k$) is given by a formal vector field; moreover, on the space of formal multivectors, Lie bracket with this latter vector field coincides with the adjunction derivation induced by $\delta$. The $-I_k$ are simply (one-form valued) vector fields corresponding to the derivation $\delta = d_k$.

Thus,
$$\nabla = d_1 - \{I^\nabla,\cdot\},$$
where $I^\nabla = \sum_{k \neq 1}I^k$. Then since $\nabla$ is flat,
\begin{align*}
  0 & = \nabla^2 \\
  &= d_1^2 - d_1\{I^\nabla,\cdot\} + \{I^\nabla, \{I^\nabla,\cdot\}\}\\
  &= d_1^2 + \{d_1 I^\nabla,\cdot\} - \frac{1}{2}\{\{I^\nabla,I^\nabla\},\cdot\}.
\end{align*}
In the last line, we pick up minus signs since $I^\nabla$ is valued in $\Omega^1(X)$, and the latter space has odd degree.

Observe that $d_1^2$ is the Riemann curvature tensor $R = R_{ij}$, viewed as a two-form on $X$ valued in endomorphisms of $T^*X$ (or $TX$ by adjunction). Here, we think of $j$ and $i$ as in the input and output indices, respectively, of $R$ as an endomorphism. Thus, from $R$, we can form the degree one action functional
$$S_R = \int_\Sigma d^2x R_{ij}\pi^i(x)\check\pi^j(x)$$
such that $\{S_R,\cdots\} = d_1^2$ (the degree of $R$ is two since it is a two-form while $\check\pi^i$, being a coordinate function of an anti-field, has degree minus one).

So consider the total function
\begin{align*}
  S &= S[g] + S[\rho] + S_\fg - I^\nabla \\
  &=: S^0 - I^\nabla,
\end{align*}
which encodes both the $\nabla$-invariance and $\g$-symmetry. Then
\begin{align*}
  d_1S + \frac{1}{2}\{S,S\} &= \frac{1}{2}\{S^0,S^0\} + \Big(d_1S^0 - \{I^\nabla, S^0\}\Big) + \Big(-d_1I^\nabla + \frac{1}{2}\{I^\nabla,I^\nabla\}\Big)\\
  &= S_R,
\end{align*}
which captures the following: $S^0$ satisfies its own the classical master equation at every point $p \in X$ due to $\g$-symmetry, $S^0$ is $\nabla$-flat, and $\nabla$ is flat.

\begin{Definition}
  We call
\begin{equation}
  d_1S + \frac{1}{2}\{S, S\} - S_R = 0. \label{globalCME}
\end{equation}
the \textit{global classical master equation}.
\end{Definition}

Write
$$S = S_{\mr{kin}} - I$$
where now $I$ captures the interactions from $S^0$ and the additional term $I^\nabla$. The global classical master equation can be reexpressed as
\begin{equation}
\Big(Q + \frac{1}{2}\{I,\cdot\} + d_1 - \frac{S_R}{\hbar}\Big)e^{I/\hbar} = 0. \label{globalCME2}
\end{equation}
From (\ref{globalCME2}), obtaining a corresponding global quantum master equation (at scale $L$) involves some complication due to the fact that $d_1^2 \neq 0$. Indeed, the naive guess that $(Q + + \hbar\div_L +d_1 - \frac{S_R}{\hbar})$ should be the quantum BV differential fails since it does not square to zero.

Following \cite{LiLi}, for $L > 0$, define the operator
$$Q_L = Q + R \circ P(0,L)$$
Here, $P(0,L)$ is the degree $-1$ derivation on functionals induced by adjunction from the map $$P(0,L): \Gamma(\Sigma, T_p^*X[-1]) \to \Gamma(\Sigma; T_pX)$$
which maps an anti-field to a field. That is, we regard the $e_i \otimes e_i$ tensor in (\ref{prop}) as being an element of $\mr{Hom}(T_p^*X[-1], T_pX)$ instead of $\Hom(T_pX, T_pX)$.
As above, $R$ is $2$-form valued endomorphism on fields given by the Riemann curvature tensor.

Thus, $Q_L$ is a degree one derivation on the space of functionals.

\begin{Lemma}
  We have
  \begin{equation}
    \left(Q_L + \hbar\div_L + d_1 - \frac{S_R}{\hbar}\right)^2 = 0. \label{Q^2}
  \end{equation}
  Furthermore, we have the following compatibility with changes of scale:
     \begin{equation}
       e^{\pd_{\hbar P(\eps,L)}}\left(Q_\eps + \hbar\div_\eps + d_1 - \frac{S_R}{\hbar}\right) = \left(Q_L + \hbar\div_L +d_1 - \frac{S_R}{\hbar}\right)e^{\pd_{\hbar P(\eps,L)}}. \label{cos}
     \end{equation}
\end{Lemma}

\Proof We have
\begin{align}
  \left(Q_L  + \hbar\div_L + d_1 - \frac{S_R}{\hbar}\right)^2 &= (Q_L + \hbar\div_L)^2 + \Big(d_1 -\frac{S_R}{\hbar}\Big)^2 + \left[Q_L + \hbar \div_L, d_1 - \frac{S_R}{\hbar}\right] \label{3terms}
\end{align}
Note that we are using commutators in the graded sense. The first term of (\ref{3terms}) is
$$(Q + \hbar\div_L)^2 + (R \circ P(0,L))^2 + [Q + \hbar \div_L, R \circ P(0,L)].$$
The first term is zero as computed previously and the second term is zero since it's the square of a degree one map. We have $[\hbar \div_L, R \circ P(0,L)] = \hbar R \circ [\div_L, P(0,L)] = 0$ since $P(0,L)$ commutes with $K_L$ as operators (the former maps anti-fields to fields while the latter maps fields to fields and anti-fields to anti-fields). Next, $[Q, R\circ P(0,L)] = R\circ[Q, P(0,L)] = R \circ K_L - R$. Thus altogether, the first term of (\ref{3terms}) is $R \circ K_L - R$.

Next, the second term of (\ref{3terms}) is $d_1^2 = R$. Indeed, $S_R^2$ squares to zero, since it's multiplication by an odd function, while the cross term vanishes since $d_1R = 0$ due to the Bianchi identity.

Finally, for the third term of (\ref{3terms}). We have
$$[Q_L, S_R] = [Q,S_R] + [R \circ P(0,L), S_R].$$
The first term equals $\int d^2x \pi^i(x)(R\pd_\mu^2 \pi)^i(x)$ which vanishes since $R$ is skew-symmetric. The second term equals
$$\int d^2xd^2x' (R^2\check{\pi})^i(x) P(0,L)(x,x')\check{\pi}^i(x') = 0$$
which vanishes since $R^2$ is symmetric and the $\check\pi$ are odd. The only non-vanishing commutator in (\ref{3terms}) is thus
$$-[\mr{div}_L, S_R] = -R \circ K_L$$
since $R$ is trace-free. We also have that the commutator involving $d_1$ vanishes, where it suffices to check $[d_1, P] = 0$. Since $d_1$ is the Levi-Civita connection, it is compatible with the isomorphism $T^*X \cong TX$ induced by the Riemannian metric on $X$, and so annihilates the identity tensor on $X$ regarded as an element of $TX \otimes TX$. This implies $[d_1,P]=0$.

Altogether, the sum of all terms of (\ref{3terms}) evaluates to $(R \circ K_L - R) + R  - R \circ K_L = 0$. Finally (\ref{cos}) follows from $-[\pd_{\hbar P(\eps,L)}, S_R/\hbar] = R \circ P(\eps,L)$, where we used that $R_{ij}$ is a skew-symmetric endomorphism.\End

Thus, we make the following definition:

\begin{Definition}The \textit{global quantum master equation} is the equation
$$\left(Q_L + d_1 + \hbar \div_L - \frac{S_R}{\hbar}\right)e^{I[L]/\hbar} = 0.$$
\end{Definition}

As before we obtain a leading order obstruction $O[L]$ which measures the failure to solve the quantum master equation, namely
$$\left(Q_L + d_1 + \hbar \div_L + \frac{S_R}{\hbar}\right)e^{I[L]/\hbar} = \hbar^n O[L]e^{I[L]/\hbar} \mod \left(\hbar^{n+1}e^{I[L]/\hbar}\right).$$
We also have $O = \lim_{L \to 0} O[L]$ exists as a local action functional of degree one, by Theorem \ref{ThmO}. Moreover, we have

\begin{Lemma}
  The functional $O$ is $d_1 + \{S,\cdot\}$ closed.
\end{Lemma}

The proof proceeds exactly the same as in the proof of Theorem \ref{ThmO}.\\

\noindent\textit{Proof of Theorem \ref{MainThm}.} By the above lemma, it follows that we need to compute the cohomology of $d_1 + \{S,\cdot\} = \{S^0,\cdot\} + \nabla$ on the global complex
\begin{equation}
  \C^* := \Omega^*(X) \otimes_{C^{\infty}(X)} C^*(\g, \Jet(\T)) \label{eq:globalcomplex}
\end{equation}
where
$$\T = \Sym^2(T^*X) \oplus T[1]X.$$
We want to compute the cohomology of $\C^*$ in degrees $0$ and $1$, corresponding to the deformation and obstruction space to global quantization, order-by-order in $\hbar$.

For a general target $X$ with $\g = 0$, we have $\C^*= \Omega^*(\Jet(\T))$. Since $\{S^0,\cdot\}$ and $\nabla$ commute (and $\nabla$ preserves ghost and anti-field degrees), we can take cohomology with respect to $\nabla$ first, i.e., we filter our complex by anti-field number. By Proposition \ref{PropJet}, this yields the space of smooth sections of $\T$ in cohomological degree zero. In particular, there is no cohomology in degree greater than zero. Thus, the diffeomorphism covariance provided by $\nabla$ is not anomalous. We get as cohomology in degree zero, the space of globally defined sections of $\Sym^2(T^*X)$ modulo the image of $\{S^0,\cdot\}$, corresponding to the fact that order by order in $\hbar$, one is entitled to modify the action by an arbitary metric on $X$ modulo infinitesimal diffeomorphisms acting on the classical metric $g_{ij}$.

For the case $X = G/H$, we obtain different information depending on whether we filter by $\g$-degree or de Rham degree, i.e. take cohomology with respect to $\nabla$ or $\{S^0,\cdot\}$ first. In the former case, using Proposition \ref{PropJet} as before, the cohomology concentrates in de-Rham degree zero, and we have that $\C^*$ is quasi-isomorphic to the complex
\begin{equation}
  C^*(\g, \Gamma(\T)), \label{eq:sectionscomplex}
\end{equation}
which is the complex (\ref{eq:Jetcomplex}) but with jets at $p$ replaced with the space of global sections of $\T$. Filtering (\ref{eq:sectionscomplex}) by anti-field degree, i.e. taking cohomology with respect to the Chevalley-Eilenberg differential, we find that the cohomology of (\ref{eq:sectionscomplex}) vanishes in degree $1$ if
\begin{equation}
  H^1(\g, \Gamma(\Sym^2(T^*X))) = H^2(\g, \Gamma(T[1]X)) = 0. \label{eq:gsemisimple}
\end{equation}
If $G$ is compact, then $\Gamma(\T)$ is a completely reducible into finite dimensional irreducible $\g$-modules. Hence, if in addition $\g$ is semi-simple, (\ref{eq:gsemisimple}) holds since $H^1(\g, M) = H^2(\g, M) = 0$ for arbitrary finite-dimensional $\g$-modules $M$. In this case, we can also deduce that the zeroth cohomology of (\ref{eq:sectionscomplex}) is given by the cokernel of
\begin{equation}
  H^0(\g, T[1]X) \overset{\{S_0,\cdot\}}{\longrightarrow} H^0(\g,\Gamma(\Sym^2(T^*X))) \label{eq:globalcok}
\end{equation}
since $H^1(\g, \Gamma(T[1]X)) = 0$. The cokernel of (\ref{eq:globalcok}) is the space $\mr{Met}^G$ of $G$-invariant metrics on $X$ modulo Lie derivatives of the classical metric $g_{ij}$ by $G$-invariant vector fields.

On the other hand, we could also have filtered $\C^*$ by de-Rham degree, i.e., take cohomology with respect to $\{S^0, \cdot\}$ first. Here, the potential obstruction lies in
\begin{equation}
  \Omega^1(\mr{Met}^G) \oplus \Omega^1(H^1(\g, \Jet(T[1]X))) \oplus \Omega^2(H^0(\g,\Jet(T[1]X))), \label{eq:globalobst}
\end{equation}
corresponding to whether one can quantize the $\g$-action and $\nabla$-action in a way compatible with each other. (In the above, we regard $H^*(\g, \Jet(\T))$ as a vector bundle over $X$, whose fiber at $p$ is $H^*(\g, \Jet_p(\T))$.) In other words, when we quantize globally, counterterms are added to the metric, $\g$-action, and $\nabla$, arising as sections of
$$\Omega^0(\Sym^2(T^*X)),\; \Omega^0(C^1(\g, \Jet(T[1]X))),\; \Omega^1(\Jet(T[1]X))$$
respectively. When we try to make these counterterms compatible with the $\g$-action (make them trivial in $\{S^0,\cdot\}$-cohomology), they may fail to be compatible with the $\nabla$-action, which appears as an obstruction belonging to (\ref{eq:globalobst}), these spaces being the target space of the differential on the $E_1$-page of the spectral sequence associated to the filtration of $\C^*$ by de Rham degree. Note we did not have an obstruction space in de Rham degree zero appearing in (\ref{eq:globalobst}) by Theorem \ref{ThmSec4}: we simply choose a strongly invariant $H_p$-quantization at every point $p \in X$.

Next, we show the vanishing of the obstruction in the last two spaces in (\ref{eq:globalobst}). By Lemma \ref{LemmaCoho}, we have
\begin{align}
  \Omega^1(H^1(\g, \Jet(T[1]X))) & \cong \Omega^1(H^1(\h_p, (T[1]X)_p)) \label{eq:obst1}\\
  \Omega^2(H^0(\g,\Jet(T[1]X))) & \cong \Omega^2(H^0(\h_p, (T[1]X)_p)) \label{eq:obst2}
\end{align}
where $p \in X$ varies. With $p$-fixed, $\h_p$ is the direct sum of a semi-simple $\h_s$ and abelian Lie algebra $\a$, and its action on $T_pX$ (which arises from the compact group $H$) is completely reducible. It follows that
$$H^*(\h_p, T_pX) = H^*(\h_s, (T_pX)_0) \otimes \Lambda^*(\a)$$
where $(T_pX)_0 \subset T_pX$ is the subspace of $\a$-invariants. The hypothesis on $\h$ now ensures that $H^i(\h_p, T_pX) = 0$ for $i = 0, 1$. As a result, the part of the obstruction to global quantization lying in (\ref{eq:obst1}-\ref{eq:obst2}) must be zero in cohomology.

Finally, we consider the space $\Omega^1(\mr{Met}^G)$. This represents an obstruction occurring due to the potential monodromy of $G$-invariant quantizations around a closed loop. When we apply $\nabla$, the $E_1$-differential, we end up with the resulting obstruction space $H^1(X, \mr{Met}^G)$. Thus, if we assume $H^1(X) = 0$, this obstruction space vanishes. (Note this last condition is true if $G$ is compact and semi-simple.)\End

\subsection{The Renormalization Group}

Finally, we mention an associated renormalization group equation that governs how our effective field theories behave under a natural action of rescaling \cite{Cos}. We have the natural scaling \begin{equation}
\begin{aligned}
  R_\ell\pi^i(x) & = \pi^i(\ell x) \\
  R_\ell\check{\pi}^i(x) & = \ell^2 \check\pi^i(\ell x), \qquad \ell > 0
\end{aligned} \label{scaling}
\end{equation}
on the space of fields and anti-fields. That is, the field has scaling dimension zero while the anti-field has scaling dimension two, which is the same as that of a $2$-form on $\Sigma$. This scaling is natural because it preserves the action, and in particular, the kinetic term of the action. This implies that the propagator, when given scaling dimension twice that of the bosonic field (and hence scaling dimension zero in two dimensions), obeys the relation
$$P(\eps,L)(\ell x, \ell y) = P(\ell^{-2}\eps,\ell^{-2}L)(x,y).$$
This implies a compatibility relation between the scaling of the effective length scale $L$ and the scaling of the fields.

Given an interaction $I$, let $R_\ell^*I$ denote $I \circ R_{\ell^{-1}}$, the interaction obtained by precomposing $I$ with an action $R_{\ell^{-1}}$ on the input fields. Observe that $R_{\ell^{-1}}$ has the affect of rescaling the length $\ell$ to unity and thus captures interactions at length scale $\ell$. Thus, given our set of effective interactions $I[L]$ encoding all renormalized Feynman diagrams at length scale $L$, consider the new set of effective interactions given by
$$\mr{RG}_\ell I[L] := R_\ell^* I[\ell^2 L],$$
changing both the length scale of the fields and the scale $L$ of the interactions.

\begin{Lemma}\cite{Cos}
  For all $\ell > 0$, then the family $\{\mr{RG}_\ell I[L]\}_L$ also constitutes a family of effective field theories. In other words, for fixed $\ell$, we have $\mr{RG}_\ell I[L'] = e^{\hbar\pd_{P(L,L')}}\mr{RG}_\ell I[L]$ for all $L,L' >0$.  Moreover, $I[L]$ satisfies the scale $L$ quantum master equation if and only if $\mr{RG}_\ell I[L]$ does as well.
\end{Lemma}

Thus, $\mr{RG}_\ell$ yields a scaling action on the space of effective field theories. In \cite{Cos}, it is shown that given a renormalization scheme, there is a bijection between effective interactions and local action functionals that agree with the classical action modulo $\hbar$. In the language of renormalized perturbation theory, such a local action functional is the renormalized action functional obtained from the bare action and a renormalization scheme. Thus, given such a renormalization scheme, $\mr{RG}_\ell$ yields a scaling action on the space of (renormalized) local action functionals. This yields for us a corresponding flow on the space of local action functionals, , the \textit{renormalization group (RG) flow}, corresponding to the infinitesimal action of the scaling $\mr{RG}_\ell$. In terms of the effective interactions, this flow is the flow given by
$$-\frac{d}{d\ell}\bigg|_{\ell=1}\mr{RG}_\ell I[L].$$
The vector field on the space of local action functionals corresponding to this flow provides the $\beta$-function. The minus sign which appears in the $\beta$-function is chosen so that the $\beta$-function provides the infinitesimal change in the renormalized local action functional that is needed to compensate for an infinitesimal change of scale while leaving the bare action unchanged.

Rather than go into the details of the general theory, we focus our attention on the renormalization group flow at one loop for our theory. Moreover, we consider only its effect on the purely bosonic part of the action, i.e., we ignore the effect on the vector fields for the $\g$-action.

We proceed following the strategy of Friedan. Given $p \in X$, the counterterms needed to renormalize the theory will involve diagrams with arbitrarily many external legs to renormalize the interactions arising from the infinite Taylor series expansion of the metric $g_{ij}$ at $p$ and the interactions $I_k$ encoding $\nabla$. Thus, a priori, the RG flow for the theory based at $p$ involves computing a flow in the space of sections of jet bundles. On the other hand, because our globally defined nonlinear sigma model has no anomaly arising from the $\nabla$-symmetry, the RG flow for the global theory, up to a cohomologically trivial term, must be $\nabla$-flat jet of a metric, i.e., it must be a flow in $\Sym^2(T^*X)$. In other words, the $\beta$-function, modulo infinitesimal diffeomorphisms of $X$, is valued in $\Sym^2(T^*X)$ and it can thus be determined by computing only the behavior of the one-loop correction to the two point function $\int_\Sigma d^2x \pd_\mu\pi^i\pd_\mu\pi^i$ at each point $p$ (a flat section of $\Jet(\Sym^2(T^*X))$ is determined by its projection to $\Jet^0(\Sym^2(T^*X))$).

It is easy to compute this one-loop correction. In normal coordinates, the first derivatives $g_{ij,k}(p)$ of the metric vanish and the second derivatives $g_{ij, k\ell}$ satisfy
$$\sum_k g_{ij,kk}(p) = \frac{2}{3}Ric_{ij}(p)$$
where $Ric_{ij}$ is the Ricci curvature. A counterterm is needed to eliminate the divergence of the regulated one-loop diagram
\begin{multline}
  -\frac{\hbar}{2}\int d^2x \sum_k g_{ij,kk}(p)\pd_\mu\pi^i(x)\pd_\mu\pi^j(x)P(\eps,L)(x,x) =\\ \hbar\left(\frac{\log \eps - \log L}{12\pi}\right)Ric_{ij}(p)\int d^2x  \pd_\mu\pi^i(x)\pd_\mu\pi^j(x). \label{2ptfun}
\end{multline}
Thus, the counterterm needed to render this diagram finite must take the form
\begin{equation}
  I^{CT}(\eps) = -\hbar\left(\frac{\log \eps}{12\pi}Ric_{ij}(p)\int d^2x  \pd_\mu\pi^i(x)\pi^j(x) + \mr{finite}\right).
\end{equation}
where $\mr{finite}$ denotes any $\eps$-independent local action functional.

From (\ref{2ptfun}), we see that a logarithmic dependence on $L$ has been introduced for the corresponding renormalized effective interaction. The infinitesimal effect of $-RG_\ell$ is to shift the corresponding renormalized interaction by an amount proportional to
$$\frac{d}{d\ell}\bigg|_{\ell=1} (\log \ell L) \int d^2x Ric_{ij}(p)\pd_\mu\pi^i(x)\pd_\mu\pi^j(x) =
\int d^2x Ric_{ij}(p)  \pd_\mu\pi^i(x)\pd_\mu\pi^j(x).$$

Based on the previous analysis, the conclusion is that an infinitesimal scaling of the effective interactions $I[L]$ at $p$ can be compensated by an infinitesimal change of the metric coupling $\Jet_p(g_{ij})$ by $\Jet_p(R_{ij})$, for every $p$. It is in this sense that the one loop renormalization group flow of the nonlinear sigma model is given by the Ricci flow
$\dot{g_{ij}} = \mr{Ric}_{ij}.$
This proves Theorem \ref{MainThm2}.

\appendix

\section{Graded Vector Spaces and Graded Manifolds}

A (real) graded vector space $V$ is an $\R$-vector space together with a decomposition $V = \oplus_{i \in \Z} V_i$ into vector spaces $V_i$ in degree $i$. An ordinary vector space yields a graded vector space concentrated in degree zero. Given a graded vector space $V$ and $n \in \Z$, the graded vector space $V[n]$ is defined by shifting the grading on $V$ via
$$(V[n])_i = V_{i + n}.$$

For ordinary vector spaces $V$, one has the familiar notion of $\Sym^n(V)$ and $\Lambda^n(V)$, the symmetric and exterior powers of $V$. For graded vector spaces, one defines symmetric powers in the graded sense. Namely, let $\otimes^n V$ be the graded vector space whose graded components are
$$(\otimes^n V)_i = \bigoplus_{i_1 + \cdots + i_n = i}V_{i_1}\otimes \cdots \otimes V_{i_n}.$$
We have an action of $\Sym_n$ such that any transposition of adjacent elements acts via
$$u \otimes v \mapsto (-1)^{|u|\cdot|v|}v \otimes u$$
where $u$ and $v$ are homogeneous elements of degree $|u|$ and $|v|$, respectively. Then $\Sym^n(V)$ is the $\Sym_n$-invariant subspace of $\otimes^n V$ with respect to the above action. We write
$$\Sym(V) = \bigoplus_{n \geq 0} \Sym^n(V)$$
to denote the total symmetric algebra on $V$. Likewise, $\Lambda^n(V)$ is the $\Sym_n$-invariant subspace of $\otimes^n V$ with respect to the action
$$u \otimes v \mapsto (-1)^{|u|\cdot|v|+1}v \otimes u.$$
We also write
$$\widehat{\Sym}(V) = \prod_{n \geq 0} \Sym^n(V)$$
to denote the completed symmetric algebra consisting of formal power series in elements of $V$.

A map $f: V \to V'$ of graded vector spaces has degree $k$ if $f(V_i) \subset V'_{i+k}$. If the degree is not specified, it is understood to be degree zero. If $f: V \to V'$ and $g: W \to W'$ are two maps, we get an induced map $f \otimes g: V \otimes W \to V' \otimes W'$ defined on homogeneous elements via $v \otimes w \mapsto (-1)^{|g||v|}f(v) \otimes g(w)$.

If $V$ is a graded vector space, then its dual space $V^*$ is the graded vector space given by
$$V^*_i = (V_{-i})^*,$$
that is, the degree $i$ component of $V^*$ is the dual space of $V_{-i}$. In this way, the evaluation pairing
$$V^* \otimes V \to \R$$
is a degree zero map.

\subsection{Directional Derivatives} \label{SecDD}

Let $\Hom(V^{\otimes n},\R)$ denote the space of $n$-multilinear maps from $V^{\otimes n}$ to $\R$. It has a natural action of $\Sym_n$ induced from the one on $V^{\otimes n}$. Let
$$\cO(V) = \wSym(V^*)$$
denote the space of formal power series functions on $V$. For $v \in V$, define the contraction operator
\begin{align}
  \pd_v: \Hom(V^{\otimes n},\R) & \to \Hom(V^{\otimes n-1},\R) \nonumber \\
v_1^* \otimes \cdots \otimes v_n^* & \mapsto \sum_i (-1)^{|v|( |v_1|+\cdots+|v_{i-1}|)}v_i^*(v) \Big(v_1^*\otimes\cdots \otimes v_{i-1}^* \otimes v_{i+1}^*\otimes\cdots \otimes v_n^*\Big). \label{eq:pd}
\end{align}
In other words, $\pd_v$ is the directional derivative with respect to $v$, where in the graded setting, it is a derivation of degree $|v|$ whence the usual sign rules apply to the Leibniz rule for $\pd_v$.

More generally, given an element of $K = u \otimes v \in V^{\otimes 2}$, we can define the operation
\begin{equation}
\pd_K = \frac{1}{2}\pd_v\pd_u. \label{eq:contractK}
\end{equation}
This operation extends bilinearly to any $K \in V^{\otimes 2}$ and is $\Sym_2$-invariant. Hence, we have a well-defined contraction operator $\pd_K$ for any $K \in \Sym^2(V)$.

\subsection{Graded Manifolds} \label{SecGM}

We give a compressed treatment here. For a more detailed treatment, see \cite{DM} where the analysis on supermanifolds generalizes readily to the graded setting. A graded manifold $(M,A)$ is a smooth manifold $M$ equipped with a sheaf of graded $C^\infty(M)$-algebras $A$. A natural way in which graded manifolds arise is by specifying a graded vector bundle $E$ over $M$ and letting $A$ be the sheaf of sections of $\Sym(E^*)$, the symmetric algebra on the dual bundle $E^*$ of $E$. (One might also want to define $A$ using the completed symmetric algebra instead). In this case, we encode this data succinctly by saying that $E$ is a graded manifold, and we regard $A$ as the sheaf of functions on $E$, which we denote by $\cO(E)$. The sheaf of derivations on $\cO(E)$ is itself a sheaf of $A$-algebras, which we call the tangent sheaf of $E$. The $A$-linear dual of the tangent sheaf is the cotangent sheaf, and we can consider the corresponding exterior algebra over $A$ of such a sheaf to obtain the sheaf of differential forms on $E$. See \cite{DM} for further reading.

The most relevant example for this paper is the shifted cotangent bundle $E = T^*[-1]M$, whose fiber above a point $p \in M$ is the cotangent space $T_p^*M[-1]$ at $p$ shifted to be in degree one. In this case, the sheaf of functions on $T^*[-1]M$ is
$$\cO(T^*[-1]M) = \Sym(T[1]M),$$
which is the space of multivector fields on $M$ with grading given by the negative of the rank of a multivector. The ordinary cotangent bundle $T^*M$ comes equipped with a canonical symplectic form, which in local Darboux coordinates $(x^i,\xi_i)$ is given by $\sum_i dx^i \wedge d\xi_i$ (the $x^i$ are local coordinates for $M$ and the $\xi_i$ are components of a cotangent vector with respect to the basis $dx^i$). For the shifted cotangent bundle $T^*[-1]M$, we automatically get an induced symplectic form, a nondegenerate closed two-form on $T^*[-1]M$. In local Darboux coordinates, it is given by the same expression as before, only now the $\xi_i$ are odd. Since the odd fibers of $T^*[-1]M$ have degree one, the coordinate functions $\xi_i$ have degree minus one and so the symplectic form on $T^*[-1]M$ has degree minus one.

\subsection{The Chevalley-Eilenberg Cochain Complex} \label{sec:CE}

Let $\fg$ be a Lie algebra over the field $k$. A $\fg$-module $M$ is simply a representation of $\fg$. The Chevalley-Eilenberg cochain complex $C^*(\fg,M)$ of $\fg$ with coefficients in $M$ is typically defined as the space $\Lambda^*(\fg, M)$, the exterior algebra on $\fg^*$ with coefficients in $M$, equipped with the Chevalley-Eilenberg differential
\begin{align*}
  d_{CE}: \Lambda^p(\fg^*, M) & \to \Lambda^{p+1}(\fg^*, M) \\
  d_{CE}\omega(Z_0,\ldots, Z_p) & = \sum_i (-1)^i Z_i\cdot\omega(Z_0,\ldots,\hat{Z_i},\ldots,Z_p) \\
   & \qquad + \sum_{i<j}(-1)^{i+j}\omega([Z_i,Z_j],Z_0,\ldots,\hat{Z_i},\ldots,\hat{Z_j},\ldots,Z_p).
\end{align*}
(In the conventions of this paper, we use the negative of the above formula for our Chevalley-Eilenberg differential.) On the other hand, if we wish $\Lambda^*(\fg^*, M)$ to be a graded vector space in which $p$-forms have grading degree $p$, then we can describe the underlying vector space for the Chevalley-Eilenberg cochain complex as
$$\Sym(\fg^*[-1]) \otimes M.$$
Observe that the natural $G$ action on $C^*(\fg, M)$ arising from the $G$ action on $M$ and the coadjoint action on $\fg^*$ commutes with the differential $d_{CE}$.

\section{Jet Bundles}\label{SecJet}

Jet bundles provide a natural geometric framework for the formal theory of differential equations. We provide a self-contained introduction here, see \cite{Sp} for further background. Given a smooth manifold $X$, let $\Jet^k_p(X)$ denote the vector space of germs at $p$ of smooth functions modulo those germs which vanish to order $k$ at $p$. In local coordinates $y^i$ on $X$, a function $u$ yields a local section $\left(\pd_{y^I} u/I!\right)_{|I| \leq k}$ of $\Jet^k(X)$. The spaces $\Jet^k_p(X)$ as $p$ varies form a smooth vector bundle $\Jet^k(X)$ over $X$ known as the bundle of $k$-jets.

The natural projections $\Jet^{k+1}_p(X) \to \Jet^k_p(X)$ allow us to define the limit
$$\Jet_p(X) = \underset{\longleftarrow}{\lim}\; \Jet^k_p(X),$$
the $\infty$-jet space at $p$ (or simply just the jet space at $p$). These spaces also fit together to form a Fr\'echet bundle over $X$. As we are interested in only the algebraic aspects of these jet bundles, we only regard the jet bundles as locally free $C^\infty(X)$ modules.

There is a natural map $\mr{jet}^k$ from smooth functions on $X$ to $\Jet^k(X)$ which assigns at every point of $p$ the corresponding $k$-jet of the section at $p$. This induces a map $\mr{jet}: C^\infty(X) \to \Jet(X)$ by assigning to every point $p$ the $\infty$-jet of the section at $p$.

The above constructions apply equally well to sections of a vector bundle $E$ instead of functions. We thus have $\Jet(E)$ the bundle of jets of smooth sections of $E$ and a map $\mr{jet}: E \to \Jet(E)$ which assigns to a smooth section of $E$ its jet. Jet bundles can be defined globally in a sheaf-theoretic fashion without the use of coordinates. We will do so later, since it will be more convenient to describe the flat connection on $\Jet(E)$ in a coordinate independent manner.

Partial differential operators between two vector bundles can be expressed as linear bundle maps between jet bundles. Namely, we have
\begin{Lemma} \label{LemmaProlong}
  A $k$th order differential operator $D: E \to F$ uniquely determines a linear map $\Jet^k(E) \to F$ and vice versa.
\end{Lemma}

\noindent\textbf{Notation.} As is standard, given a vector bundle $E$, by abuse of notation we will often denote its sheaf of smooth sections by the same letter. If we wish to be more precise, we will denote the latter by $C^\infty(E)$ or $\Gamma(E)$.\\

Lemma \ref{LemmaProlong} is essentially tautologous since a $k$th order differential operator is locally a linear combination of derivatives up to order $k$. For every $\ell \geq 0$, the map $\mr{jet}^\ell: F \to \Jet^\ell(F)$ is an $\ell$th order differential operator. Thus, by the above lemma, the composition $\mr{jet}^\ell \circ D$ induces a map $\Jet^{k+\ell}(E) \to \Jet^\ell(F)$. Since these maps are all compatible for $\ell \geq 0$, then $D$ induces a linear map
$$p(D): \Jet(E) \to \Jet(F)$$
on the jet bundle. It is called the \textit{prolongation} of $D$. Concretely, the prolongation of a differential operator is obtained by successively differentiating the equation $Du = 0$ and encoding the resulting infinite family of differential equations into a single linear operator on the jet bundle.

The bundle $\Jet(E)$ was defined as a limit of bundles $\Jet^k(E)$. The dual bundle
$$\Jet(E)^* = \mr{Hom}(\Jet(E),C^\infty(X))$$
is defined as the colimit of bundles $\mr{Hom}({\Jet}^k(E), C^\infty(X))$. In simple terms, an element of $\mr{Hom}(\Jet^k(E),C^\infty(X))$ is an order $k$-differential operator from $E$ to the trivial bundle by Lemma \ref{LemmaProlong}. Thus, $\Jet(E)^*$ is the space of all differential operators from $E$ to $\R$, that is, it is the free $D_X$-module
$$D_X \otimes_{C^\infty(X)} E^*.$$
Here $D_X$ stands for the $C^\infty(X)$-sheaf of differential operators on $X$.

A \textit{polydifferential function} on $E$ is a function which is locally a polynomial in the derivatives of sections of $E$. In other words, it is an element of $\Sym_{C^\infty(X)}(\Jet(E)^*)$. A formal polydifferential function is then an element the completed ring
$$\mr{PolyDiff}(E) = \wSym_{C^\infty(E)}(\Jet(E)^*),$$
a formal power series in polydifferential operators. For brevity, we will not distinguish between formal and ordinary polydifferential operators. Note that $\mr{PolyDiff}(E)$ is again a $D_X$-module in the natural way obtained from applying the Leibniz rule. Namely, if $D \in D_X$ is a differential operator and $D_1(f)\cdots D_N(f)$ is a polydifferential function of a section $f$ of $E$, with $D_i$ differential operators, then
$$D(D_1(f)\cdots D_N(f)) = (DD_1)(f)\cdots D_N(f) + \ldots + D_1(f)\cdots (DD_N)(f).$$
As usual, sign rules must be applied when applying the Leibniz rule if any of these objects are graded.

A local action functional on the space of sections of $E$ is given by integrating a polydifferential function of $E$ against a density on $X$. Because we are only interested in densities modulo those which are exact, this means that the space of local action functionals is given by
$$\cO_\loc(E) = \Dens(X) \otimes_{D_X} \mr{PolyDiff}(E).$$
Here, $\Dens(X)$ is the sheaf of densities on $X$ regarded as the dual $C^\infty(X)$-sheaf to the sheaf of compactly supported smooth functions on $X$. It is a right $D_X$-module in the natural way. Namely, given $\omega \in \Dens(X)$, $D \in D_X$, and $f \in C^\infty_0(X)$, we have $\int_X (\omega D)f = \int_X \omega (Df).$

Given a local action functional $S$, the \textit{variational derivative} $\delta S$ of $S$ is a local one-form on the space of local action functionals, i.e. an element of
$$\Dens(X) \otimes_{C^\infty(X)} (E^* \otimes_{C^\infty(X)} \mr{PolyDiff}(E)).$$
It is defined as follows. We have that $E$ acts as a derivation in the natural way on $\mr{PolyDiff}(E)$, since we have the map
$$E \overset{\mr{jet}}{\longrightarrow} \Jet(E) \subset \Hom(\Jet(E)^*, C^\infty(X)),$$
where the latter inclusion comes from the natural evaluation pairing. Thus, for a section $v$ of $E$, we have $\pd_{\mr{jet}(v)} : \mr{PolyDiff}(E)) \to \mr{PolyDiff}(E))$ as given by the formula (\ref{eq:pd}) adapted to $C^\infty(X)$-sheaves. Thus, given an element $\frak{S} \in \mr{PolyDiff}(E))$, it yields the element $\delta \frak{S} \in E^* \otimes_{C^\infty(X)} \mr{PolyDiff}(E)$ which maps $v \in E$ to $\pd_v \frak{S}$. The operation $\delta$ carries over when we tensor with densities, from which we obtain the variational derivative of a local action functional.

\subsection{Coordinate Systems}

A coordinate system $\Theta_p: T_pX \to X$ (defined in a neighborhood of the origin of $T_pX$) yields an algebra isomorphism $\Jet_p(X) \cong \wSym(T_p^*X)$. Namely given a smooth function $f$, its germ at $p$ determines the Taylor series of $\Theta_p^*f: T_pX \to \R$ at the origin. Likewise, given a natural tensor bundle $\T$ over $X$, with fiber $\T_p$ over $p$, we have $\Jet_p(\T) \cong \wSym^*(T_p^*X) \otimes_\R \T_p$, where given a tensor $T$, we consider the Taylor series of the components of $\Theta_p^*(T)$ with respect to the coordinate basis tensors of $T_pX$.

In fact a coordinate system at $p$ is really the data of an isomorphism $\Jet_p(X) \cong \wSym(T_p^*X)$ along with a choice of basis for $T_pX$ (the choice of an ordered set of coordinates). Since as $p$ varies all points of $X$, $TX$ is not in general globally trivializable, all that is desired is just an isomorphism $\Jet_p(X) \cong \wSym(T_p^*X)$. This can be regarded as a choice of affine coordinate system at $p$ (a coordinate system modulo the natural action of translations and $GL_n(T_pX)$ induced from the action on $T_pX$). A global choice of affine coordinate systems is then a choice of isomorphism $\Jet(X) \cong \wSym(T^*X)$. Observe that any such isomorphism (of algebras) is determined by a splitting of the natural projection
$$F^1\Jet(X) \to F^1\Jet(X)/F^2\Jet(X) \cong T^*X,$$
where $F^k\Jet(X)$ is the subbundle of $\Jet(X)$ consisting of those jets that vanish to order $k$ at every point. Such a splitting always exists and the space of all such splittings is contractible.

Given any vector bundle $E$, then $\Jet(E)$ has a natural flat connection $\nabla: \Jet(E) \to \Omega^1(\Jet(E))$. This can be described succinctly and in a coordinate invariant way as follows. Let $Y = X$ and consider the projections $\pi_X$ and $\pi_Y$ of $X \times Y$ onto the first and second factor, respectively. Let $I$ denote the ideal in $C^\infty(X\times Y)$ consisting of smooth functions that vanish along the diagonal and let $\pi_Y^*C^\infty(X)$ denote $C^\infty(X \times Y)$ regarded as a $C^\infty(X)$-sheaf in the obvious way. Then as $C^\infty(X)$ sheaves, we have
$$\Jet^k(X) = (\pi_X)_*(\pi_Y^*(C^\infty(X))/I^k).$$
Given a function $f$, then $\jet^k(f)$ is given by the equivalence class of $\pi_Y^*(f)$ in $\Jet^k(X)$.

Let $d$ denote the de Rham differential with respect to $X$. We have the corresponding deRham complex $\Omega^*(X) \otimes_{C^\infty(X)} C^\infty(X \times Y)$. Tensoring with $\pi_Y^*(E)$, taking the quotient by $I^k$, and pushing forward to $X$, we obtain a de Rham complex
$$\Omega^0(\Jet^k(E)) \overset{d}{\longrightarrow} \Omega^1(\Jet^{k-1}(E)) \cdots \overset{d}{\longrightarrow} \Omega^n(\Jet^{k-n}(E))$$
for $k \geq n = \mr{dim}(X)$.

\begin{Proposition}\cite[Prop 1.3.2]{Sp} \label{PropJet}
  The sequence
  \begin{equation}E \overset{\mr{jet}^k}{\longrightarrow} \Omega^0(\Jet^k(E)) \overset{d}{\longrightarrow} \Omega^1(\Jet^{k-1}(E)) \cdots \overset{d}{\longrightarrow} \Omega^n(\Jet^{k-n}(E)) \to 0 \label{eq:deRham}
  \end{equation}
  is exact.
\end{Proposition}

Note that in \cite{Sp}, it is shown that (\ref{eq:deRham}) is exact as a sheaf of $C^\infty(X)$-modules. However, since the space of smooth sections of $E$ is a fine sheaf, this implies the stronger statement that (\ref{eq:deRham}) is exact as a complex of vector spaces.

Taking limits, we obtain the exact sequence
 \begin{equation}
 E \overset{\mr{jet}}{\longrightarrow} \Omega^0(\Jet(E)) \overset{d}{\longrightarrow} \Omega^1(\Jet(E)) \cdots \overset{d}{\longrightarrow} \Omega^n(\Jet(E)) \to 0 \label{eq:exactjet}
  \end{equation}
The map from $d: \Omega^0(\Jet(E)) = \Jet(E) \to \Omega^1(\Jet(E))$ is the flat connection on $\Jet(E)$. For another approach to defining this flat connection, see \cite{CFT}. Exactness of (\ref{eq:exactjet}) implies that its flat sections consist of those that are in the image of $\jet$.

\section{Wick's Theorem}

Wick's Theorem gives a combinatorial formula for the integration of monomials against Gaussian measures. We assume the reader is familiar with this lemma and only record it here for notational purposes. Consider $\R^d$ with the standard monomial basis $x^1, \ldots, x^d$ and let $A = A_{ij}$ be a symmetric nondegenerate $d\times d$ matrix. It determines a bilinear form $(x,Ax) := A_{ij}x^ix^j$ and a normalized Gaussian measure $d\mu_A = \frac{\sqrt{\det A}}{(2\pi)^{d/2}}e^{-(x,Ax)/2}d^dx$.

\begin{Lemma}(Wick's Theorem)
Consider the monomial $f(x) = x^{i_1}\cdots x^{i_n}$. Then
\begin{equation}
  \int f(x)d\mu_A = \frac{1}{2^{n/2}(n/2)!}\sum_{\sigma \in S_n}A^{i_{\sigma(1)}i_{\sigma(2)}}\cdots A^{i_{\sigma(n-1)}i_{\sigma(n)}}.
\end{equation}
for $n$ even and is zero otherwise. Here $A^{ij}$ denotes the inverse matrix of $A_{ij}$.
\end{Lemma}
As is well-known, the sum on the right-hand-side has an elegant description in terms of Feynman diagrams.

We can encode Wick's Theorem more succinctly as follows. Let $W$ be a finite dimensional Euclidean vector space of dimension $d$. The nondegenerate bilinear form $(\cdot,A\cdot)$ on $W$ determines an isomorphism $W \cong W^*$. This isomorphism allows us to transfer the bilinear form $(\cdot,A\cdot)$ on $W$ to a bilinear form on $W^*$. Thus, if $P \in \Sym^2(W)$ is the two-tensor which induces this latter bilinear form, then $\pd_P$ is the operation of Wick contraction when integrating against $d\mu_A$. It is then easy to see that the operation
$$e^{\pd_P} = \sum_{n=0}^\infty \frac{(\pd_P)^n}{n!}$$
is precisely the sum over all possible choices of Wick contractions (with all the correct combinatorial factors accounted) using $n$ propagators $P$, with $n$ ranging over all possible values. In particular, we have the following:

\begin{Lemma}(Wick's Theorem, second version)
  Let $P$ be as above. Then
  \begin{equation}
      \int f(x)d\mu_A = (e^{\pd_P}f)(0).
  \end{equation}
\end{Lemma}
Here, the right-hand side is evaluated at zero so that the maximal number of Wick contractions have been made (i.e., in the Feynman-diagrammatic picture, we only sum over vacuum diagrams).

Returning to the general situation, consider now the formal power series $e^{I/\hbar}$, where $I \in \widehat{\Sym}(W^*)[[\hbar]]$ is a power series function on $W$ and $\hbar$ is a formal parameter. Suppose we want to generate all Feynman diagrams with vertices labelled by interactions from $I$ and edges labelled by the propagator $P$. Then the general Feynman diagrammatic combinatorics implies that the sum over all Feynman diagrams equals the exponential of only those which are connected\footnote{Connected, recall, means no component of the graph is a vacuum diagram, i.e., a diagram with no external tails.}:

\begin{Lemma} \label{LemmaFD} Let $I$ be at least cubic modulo $\hbar$. Then
  \begin{equation}
    e^{\hbar \pd_P}e^{I/\hbar} = \mr{exp}\left(\hbar^{-1}\sum_{\gm  \;\mr{ connected}} I_\gm\right)
  \end{equation}
where $I_\gm \in \Sym(W^*)[[\hbar]]$ is the interaction corresponding to the connected Feynman diagram $\gamma$ (with the appropriate combinatorial factor). The sum in the exponent is understood to be valued in $\widehat{\Sym}(W^*)[[\hbar]]$, the space of formal power series on $W$ valued in formal power series in $\hbar$.
\end{Lemma}

The advantage of this notation is that it succinctly packages all the algebraic manipulations involving Wick's Theorem. Moreover, since
$$e^{\pd_P}e^{\pd_{P'}} = e^{\pd_{P+P'}}$$
it follows that $e^{\pd_P}$ is invertible with inverse $e^{-\pd_P}$.

\section{Effective Field Theories}

Our goal here is to motivate the length scale based approach to effective field theories used in this paper. In this approach, a length parameter enters as a regularizing parameter in the heat-kernel method of regularization. Because this approach is less intuitive than an approach based on energy scales, we first explain the latter and then see how it adapts to a length scale formulation. 

\subsection{Energy Scale Approach}

We begin by working in a finite dimensional setting. Let $W$ be a finite dimensional Euclidean vector space with inner product $(\cdot,\cdot)$ and let $d\mu$ denote the associated Lebesgue measure. Let $Q$ be a nonpositive self-adjoint operator. We then obtain the ``Gaussian" measure
$$d\mu_Q = ce^{(\phi, Q\phi)/2\hbar}d\mu(\phi),$$
where $c$ is a constant chosen so that $d\mu_Q$ is a normalized Gaussian measure when restricted to the span of the eigenspaces on which $Q$ is nondegenerate.

Let $W_\lambda$ denote the $\lambda$ eigenspace of $-Q$. Given $\Lambda \geq 0$, we introduce the following objects:
\begin{align*}
  W^>_\Ld &= \oplus_{\lambda > \Lambda}W_\lambda \\
  W^{\leq}_\Ld &= \oplus_{\ld \leq \Ld}W_\ld \\
  d\mu^>_{Q,\Ld} &= d\mu_Q|_{W^>_\Ld} \\
  d\mu_{Q,\Ld}^{\leq} &= d\mu_Q|_{W^\leq_\Ld}
\end{align*}
For brevity, we also write $d\mu_{Q,\Ld}$ for $d\mu_{Q,\Ld}^\leq$.

Let $\phi^a_\lambda$ denote an orthonormal eigenbasis for $Q$, where $-Q\phi^a_\lambda = \lambda \phi^a_\lambda$. Given $0 \leq \Lambda_1 < \Lambda_2$, we have the associated propagator
$$P(\Lambda_1,\Lambda_2) = \sum_{\Lambda_1 < \lambda < \Lambda_2}\frac{1}{\lambda}\phi^a_\lambda \otimes \phi^a_\lambda,$$
whose corresponding operator
$$\sum_{\Lambda_1 < \lambda < \Lambda_2}\frac{1}{\lambda}\phi^a_\lambda (\phi^a_\lambda,\cdot)$$
is an approximate inverse to $-Q$. In fact, it is the inverse of $-Q$ when restricted to $\oplus_{\Ld_1 < \ld < \Ld_2}W_\ld$.

Let $V$ be a polynomial vector field on $W$, i.e., $V \in \Sym(W^*)\otimes W$.

\begin{Lemma}
  We have the identity %\red{up to normalization}
  \begin{equation}
    \int_{W^>_\Ld} \L_V d\mu_Q = \L_{V[\Ld]}d\mu_{Q,\Lambda}
  \end{equation}
  where $V[\Ld] = e^{\hbar \pd_{P(\Ld,\infty)}}V|_{W^\leq_\Ld}$ is a vector field on $W^\leq_\Ld$.
\end{Lemma}

\Proof On top degree differential forms, we have that $\L_V = d\iota_V$, where $\iota_V$ denotes contraction with the vector field $V$ and $d$ is exterior derivative. Write $d = d_> + d_\leq$ to denote the components of $d$ belonging to the $W^>_\Ld$ and $W^\leq_\Ld$ directions, respectively. Then
\begin{align*}
  \int_{W^>_\Ld} \L_V d\mu_Q &= \int_{W^>_\Ld}(d_> + d_\leq)\iota_V d\mu_Q \\
  &= \int_{W^>_\Ld}d_\leq\iota_V(d\mu^>_{Q,\Lambda}d\mu_{Q,\Lambda}) \\
  &= d_\leq \int_{W^>_\Ld}d\mu^>_{Q,\Lambda}(\iota_Vd\mu_{Q,\Lambda})\\
  &= d_\leq \iota_{V[\Lambda]}d\mu_{Q,\Lambda}.
\end{align*}
In the second line, the term arising from $d_>$ vanishes since it is exact. Likewise, in passing to line three, the only nonzero contribution to the integral arises from $V$ contracting with the $d\mu_{Q,\Ld}$ factor. Finally, the last line is an application of Wick's Theorem.\End

Let $QV$ denote $\L_V (\cdot, Q\cdot)/2$. Then what the above shows is that
$$\L_Vd\mu_Q = \left(\frac{QV}{\hbar} + \mr{div}V\right)d\mu_Q = 0$$
implies
$$\L_{V[\Ld]}d\mu^\Ld_Q = \left(\frac{QV[\Ld]}{\hbar} + \div_\Lambda V[\Lambda]\right)d\mu^\Ld_Q = 0$$
where $\mr{div}$ and $\mr{div}_\Lambda$ are the divergence operators with respect to the Lebesgue measures $d\mu$ and $d\mu|_{W^{\leq}_\Ld}$, respectively. In other words, the invariance of the measure $d\mu_Q$ with respect to the vector field $V$ descends to an invariance of the measure $d\mu_Q^\Lambda$ with respect to the scale $\Lambda$ vector field $V[\Ld]$.

We can generalize the above situation to formal (i.e. defined as a power series) non-Gaussian measures as follows. Observe that $\L_V (fd\mu_Q) = \L_{fV}d\mu_Q$ for any function $f$. In particular, if we wish to consider an interaction term $f = e^{I/\hbar}$, with $I \in \widehat{\Sym}(W^*)$, we can repeat the preceding analysis but with $V$ replaced with $e^{I/\hbar} V$. One can verify that
\begin{equation}
  (fV)[\Lambda] = e^{I[\Lambda]/\hbar}V_I[\Lambda] \label{fV}
\end{equation}
where $I[\Lambda] \in \widehat\Sym(W^*)[[\hbar]]$ is given by
\begin{equation}
  e^{\hbar \pd_{P(\Ld,\infty)}}e^{I/\hbar}|_{W^\leq_\Lambda} =: e^{I[\Lambda]/\hbar} \label{ILambda}
\end{equation}
and
$$V_I[\Lambda] = e^{-I[\Lambda]/\hbar}e^{\hbar\pd_{P(\Ld,\infty)}}(e^{I/\hbar}V)$$
is the sum of all connected Feynman diagrams such that exactly one vertex is labelled by $V$ and the rest are with $I/\hbar$. Thus,
$$\L_V\left(e^{I/\hbar}d\mu_Q\right) = \left(\frac{QV}{\hbar} + \frac{\L_V I}{\hbar} + \mr{div}V\right)d\mu_Q = 0$$
implies
\begin{equation}
\L_{V_I[\Ld]}\left(e^{I[\Lambda]/\hbar}d\mu^\Ld_Q\right) = \left(\frac{QV_I[\Ld]}{\hbar} + \frac{\L_{V_I[\Lambda]}I[\Lambda]}{\hbar} + \div_\Lambda V_I[\Lambda]\right)d\mu^\Ld_Q = 0. \label{QMEdeg0}
\end{equation}
This latter equation expresses the invariance of the measure $e^{I/\hbar}d\mu_Q$ with respect to the vector field $V$ in terms of objects defined at scale $\Lambda$, namely, the scale $\Lambda$ effective interactions $I[V]$ and the scale $\Lambda$ vector field $V[\Lambda]$. In other words, a symmetry  for the full theory descends to a scale $\Lambda$ symmetry for the effective theory defined at scale $\Lambda$. This equation is precisely the scale $\Lambda$ quantum master equation for this simple example. Such an equation also goes by the name Ward identity and it captures how a classical symmetry is manifest at the quantum level.

We can now derive the full quantum master equation, with the previous simplified example providing justification for why this more general quantum master equation is sensible. Consider the above (still finite-dimensional) situation in which we have a power series action functional $S_0$ on $W$ invariant under the action of a Lie algebra $\g$. As in Section \ref{SecBV}, we can extend $S_0$ to a function $S$ on $W \oplus W^*[-1] \oplus \g[1] \oplus \g^*[-2]$ that satisfies a classical master equation. Write
$$S = S_{\mr{kin}} - I$$
where $S_{\mr{kin}} = (\phi, Q\phi)/2$ is the quadratic part of $S_0$ and $I =  -(S - S_{kin})$ is the interaction term depending on $\phi$ and the other fields. Note that unlike (\ref{QMEdeg0}), where we have separated a symmetry $V$ from the interaction $I$ for the sake of illustration, in our present situation, the interaction $I$ includes all the symmetries in question. This is a convenient tactic, because then all the symmetries participate in Feynman diagrams, thereby descending to the appropriate form and receiving all necessary quantum corrections when we go to scale $\Lambda$. That is, $I[\Lambda]$ defined by
$$e^{\pd_{\hbar P(\Ld,\infty)}}e^{I/\hbar} =: e^{I[\Lambda]/\hbar},$$
is a sum of multivector fields. The \textit{scale $\Lambda$ quantum master equation} is defined to be
\begin{equation}
(Q + \mr{div}_\Lambda)e^{I[\Ld]/\hbar} = 0, \label{QMELd}
\end{equation}
where $\mr{div}_\Lambda$ is the sum of the two terms, the divergence with respect to $d\mu_\Lambda$ and the operator $\mr{div}_\g$ (as in \ref{div2terms}). When $\g = \R$ (the symmetry of the theory is generated by single vector field $V$), this recovers exactly (\ref{QMEdeg0}). When $\dim \g > 1$, the full quantum master equation (\ref{QMELd}) expresses (i) the invariance of the scale $\Lambda$ measure with respect to scale $\Lambda$ vector fields arising from each vector of $\g$; (ii) higher compatibility conditions among these scale $\Ld$ vector fields. The scale $\Lambda$ vector fields are given by the component of $I[\Lambda]$ of multivector field degree one.

\subsection{Length Scale Approach}

In the context of quantum field theory, the above picture has to be modified due to the process of regularization and renormalization. Renormalization introduces counterterms which a priori may destroy the underlying symmetries of the classical theory, in particular, the invariance of the ``quantum measure" with respect to symmetries. Thus, having a scale $\Lambda$ quantum master equation hold means that no such symmetry violation occurs for the effective theory at scale $\Lambda$.

The previous approach to defining a regulated quantum master equation was based on energy, namely, using a momentum cutoff to define the propagator. The length scale approach is based on the heat kernel regularization of the propagator, whereby for a theory in which the kinetic operator $-Q$ of the theory is a Laplace type operator $\Delta$, we take as regulated propagator
$$P(\eps,L) = \int_\eps^L e^{-t\Delta}dt.$$
Since the integral kernel of $e^{-t\Delta}$ is localized to length scales of order $t^{1/2}$, $\eps$ is a small length (ultraviolet) regulator while $L$ is a large length (infrared) regulator. The length scale approach is convenient because then effective theories $I[L]$ indexed by length become more and more local as $L \to 0$, as can be deduced from the heat kernel localizing to the diagonal at small times. In this way, one can establish Theorem \ref{ThmO} which ensures that the obstruction $O$ becomes a local action functional as $L \to 0$. In comparison, the energy scale approach is less compatible with locality, since even at high energies (small length), the operation of projection onto large eigenvalues of the Laplacian is a highly nonlocal operation.

The length scale $L$ quantum master equation is defined by formal analogy with the quantum master equation based on the energy scale $\Lambda$. Given our regulated propagator $P(\eps,L)$, we obtain a set of regulated interactions $I[\eps,L]$ via
$$e^{I[\eps,L]/\hbar} = e^{\hbar\pd_{P(\eps,L)}}e^{I/\hbar}.$$
Renormalization means that we introduce local $\eps$-dependent counterterms $I^{CT}(\eps)$ valued in a power series in $\hbar$, so that the limit
$$\lim_{\eps \to 0}e^{\hbar\pd_{P(\eps,L)}}e^{\left(I + I^{CT}(\eps)\right)/\hbar}$$
exists as a series in $\hbar$ \cite{Cos}. This defines for us a set of renormalized, effective interactions $I[L]$ at length scale $L$ given by
$$e^{I[L]/\hbar} = \lim_{\eps \to 0}e^{\hbar\pd_{P(\eps,L)}}e^{(I + I^{CT}(\eps))/\hbar}.$$
We define the \textit{(length) scale $L$ quantum master equation} by replacing all objects occurring in the energy  scale quantum master equation (\ref{QMELd}) with their length scale counterparts. Thus, $I[\Lambda]$ is replaced with $I[L]$ and $\mr{div}_\Lambda$ is replaced with $\mr{div}_L$. Hence, we obtain the equation
\begin{equation}
  (Q +\mr{div}_L)e^{I[L]/\hbar} = 0. \label{eq:QMEapp}
\end{equation}
The fact that this is a sensible equation that deserves to be called the quantum master equation arises from the fact that $(Q + \mr{div}_L)$ squares to zero and that Theorem \ref{ThmO} holds. Equation (\ref{eq:QMEapp}) expresses how a symmetry is manifest at the quantum level for an effective field theory at length scale $L$.
\pagebreak

%auto-ignore


\begin{thebibliography}{AKSZ}
\bibitem{Abb} L. F. Abbott. \textit{Introduction to the background field method.} Act. Phys. Pol. Vol B13 (1982), 33--50.
\bibitem{AGM} L. Alvarez-Gaum\'e, D. Freedman, and S. Mukhi. \textit{The background field method and the ultraviolet structure of the supersymmetric nonlinear $\sigma$-model}. Ann. Phys. 134 (1981), 85--109.
\bibitem{AKSZ} M. Alexandrov, M. Kontsevich, A. Schwarz, and O. Zabronsky. \textit{Geometry of the master equation}. Internat. J. Modern Phys. A 12 (1997), no. 7, 1405--1429.
\bibitem{AS} S. Axelrod and I. M. Singer. \textit{Chern-Simons perturbation theory. II} J. Diff. Geom.
    vol. 39, no. (1994), 173--213.
\bibitem{BV} I. Batalin and G. Vilkovisky. \textit{Gauge algebra and quantization.} Phys. Lett. B. 102, 27--31.
\bibitem{BLS} W. Bardeen, B. Lee, and R. Shrock. \textit{Phase transition in the nonlinear $\sigma$ model in a $(2+\eps)$-dimensional continuum.} Phys. Rev. D. 14 (1976), 985--1005.
\bibitem{BPS} W. A. Bardeen, O. Piguet, and K. Sibold. \textit{Mass generation in a normal-product formulation.} Phys. Lett. B., 1977, Vol. 72, No. 2,  p. 231--232.
\bibitem{BBBCD} C. Becchi, A. Blasi, G. Bonneau, R. Collina, and F. Delduc. \textit{Renormalizability and infrared finiteness of non-linear $\sigma$-models: A regularization-independent analysis for compact coset spaces.} Comm. Math. Phys. 120 (1988) 121--148.
\bibitem{BRS} C. Becchi, A. Rouet, R. Stora. \textit{Renormalization of gauge theories.} Ann. Phys. 98 (1976), 287--321.
\bibitem{LNP} P. Breitenlohner, D. Maison, and K. Sibold. \textit{Lecture Notes in Physics.} vol. 303, Springer-Verlag, Berlin. 1988.
\bibitem{BLZ} E. Brezin, J. C. Le Gillou, and J. Zinn-Justin. \textit{Renormalization of the nonlinear $\sigma$-model in $2+\eps$ dimensions.} Phys. Rev. D 14 (1976), no. 10, 2615--2621.
\bibitem{CL} D. M. Capper and G. Liebbrandt. \textit{Dimensional regularization for zero-mass particles in quantum field theory.} J. Math. Phys. 15, 82--85 (1974).
\bibitem{CFT} A. Cattaneo, G. Felder, and L. Tomassini. \textit{From local to global deformation quantization of Poisson manifolds.} Duke Math. J., vol. 115, no. 2 (2002), 329--352.
\bibitem{Col} J. C. Collins. \textit{Renormalization. An introduction to renormalization, the renormalization group, and the operator-product expansion.} Cambridge University Press, Cambridge, 1984.
\bibitem{Cos} K. Costello. \textit{Renormalization and effective field theory.} Math. Surveys and Monographs, 170. Amer. Math. Soc., Providence, RI, 2011.
\bibitem{CosWG} K. Costello. \textit{A Geometric construction of the Witten genus II.} \url{arxiv:1112.0816}
\bibitem{QFT-Math} P. Deligne, P. Etingof, et al. \textit{Quantum fields and strings: A course for mathematicians}, vol. 1, 2. Amer. Math. Soc., Providence, RI. 1999.
\bibitem{CPS} T. E. Clark, O. Piguet., and K. Sibold. \textit{Infrared anomaly induced by spontaneous breakdown of supersymmetry.} Nuc. Phys. B. vol. 119, 1977, 292--314.
\bibitem{DM} P. Deligne and J. Morgan, \textit{Classical Fields and Supersymmetry} in \textit{Quantum fields and strings: a course for mathematicians}, Vol. 1, 2 (Princeton, NJ, 1996/1997) , Amer. Math. Soc., Providence, RI, 1999.
\bibitem{FR} K. Fredenhagen and K. Rejzner. \textit{Batalin-Vilkovisky formalism in perturbative algebraic quantum field theory.} Comm. Math. Phys. 317 (2013), 697-725.
\bibitem{Fr} D. Friedan. \textit{Nonlinear sigma models in $2+\eps$ dimensions.} Ann. Phys. 163 (1985), 318--419.
\bibitem{GK2} K. Gawedzki and A. Kupiainen. \textit{Continuum limit of the hierarchical O(N) non-linear $σ$-model.} Comm. Math. Phys. 106 (1986), 533-550.
\bibitem{GJ} J. Glimm and A. Jaffe. \textit{Quantum physics: a functional integral point of view.} Springer-Verlag, New York, NY, 1987.
%\bibitem{GL} Gell-Mann and Levy. \textit{The Axial Vector Current in Beta Decay.} Nuovo Cimento 16 (1960), 705--726.
\bibitem{HT} M. Henneaux and C. Teitelboim. \textit{Quantization of gauge systems.} Princeton University Press, Princeton, NJ. 1992.
\bibitem{Hon} J. Honerkamp. \textit{Chiral Multi-Loops.} Nuc. Phys. B. 36 (1972), 130--140.
\bibitem{Ket} S. Ketov. \textit{Quantum nonlinear sigma models.} Texts and Monographs in Physics. Springer-Verlag, Berlin. 2000.
\bibitem{Lam} Y.-M. Lam. \textit{Perturbation Lagrangian theory for scalar fields -- Ward-Takahashi identity and current algebra.} Phys. Rev. D. vol. 6, no. 8 (1972), 2145--2161.
\bibitem{LiLi} Q. Li and  S. Li. \textit{On the B-twisted topological sigma model and Calabi-Yau geometry}. J. Diff. Geom.
Vol. 102, No. 3 (2016), 409--484.
\bibitem{Lucha} W. Lucha. \textit{Quadratic divergences in dimensional-regularization finite quantum field theories} Phys. Lett. B. 191 (1987), no. 4, 404--408.
\bibitem{Meetz} K. Meetz. \textit{Realization of chiral symmetry in curved isospin space.} J. Math. Phys. 10 (1969), 589--593.
\bibitem{MR} P. K. Mitter and T. R. Ramadas. \textit{The two-dimensional $O(N)$ nonlinear $\sigma$-model: renormalisation and effective actions.} Comm. Math. Phys. (1989), 575--596.
\bibitem{PR} O. Piguet and A. Rouet. \textit{Symmetries in perturbative quantum field theory.} Phys. Rep. 76, No. 1 (1981), 1--77.
\bibitem{Poly} A. M. Polyakov. \textit{Interaction of Goldstone particles in two dimensions.} Phys. Lett. B. 59, (1975), 79--81.
\bibitem{R} K. Rejzner. \textit{Batalin-Vilkovisky formalism in locally covariant field theory.} Ph.D. thesis. 2011. University of Hamburg. \url{arxiv:1111.5130}
\bibitem{Sch} A. Schwarz. \textit{Geometry of Batalin-Vilkovisky quantization.} Comm. Math. Phys. 155, 249--260 (1993).
\bibitem{Sp} D. C. Spencer. \textit{Overdetermined systems of linear partial differential equations.} Bull. Amer. Math. Soc. 75 (1969), 179--239.
\bibitem{Stelle} K. Stelle. \textit{Non-linear field renormalizations in the background field
method} in \textit{Lecture Notes in Physics.} vol. 303, Springer-Verlag, Berlin, Germany. 1988.
\bibitem{TvN} W. Troost and P. van Nieuwenhuizen and A. Van Proeyen. \textit{Anomalies and the Batalin-Vilkovisky Lagrangian Formalism.} Nuc. Phys. B 333 (1990), 727--770.
\bibitem{Wei} C. Weibel. \textit{An introduction to homological algebra.} Cambridge Studies in Advanced Mathematics, 38. Cambridge University Press, Cambridge, 1994.
\bibitem{Wein68} S. Weinberg. \textit{Nonlinear realizations of chiral symmetry.} Phys. Rev. 166 (1968), 1569--1577.
\bibitem{Wein} S. Weinberg. \textit{Quantum theory of fields. } Vol. II. Cambridge University Press, Cambridge, UK. 2005.
\bibitem{Wit-Bos} E. Witten. \textit{Non-abelian bosonization in two dimensions.} Comm. Math. Phys. 92 (1984), 455-472.
\bibitem{Wit-Hol} E. Witten. \textit{On holomorphic factorization of WZW and coset models.} Comm. Math. Phys. 144 (1992), 189--212.
\bibitem{Zim} W. Zimmerman. \textit{Composite operators in the perturbation theory of renormalizable interactions.} Ann. Phys. 77 (1973), 536--569.
\end{thebibliography}
\end{document}